\documentclass[reprint, twocolumn, showpacs, floatfix, superscriptaddress, aps, prb]{revtex4-2}
\pdfoutput=1
\usepackage{graphicx}
\usepackage{dcolumn}
\usepackage[dvipsnames]{xcolor}
\usepackage{amsmath}
\usepackage{mathrsfs}
\usepackage{physics}
\usepackage{amssymb}
\usepackage{color}
\usepackage[colorlinks, linkcolor=blue, citecolor=blue, urlcolor=blue]{hyperref}
\setcitestyle{numbers,square}

\begin{document}

\title{Generalized Gruneisen formalism with application in magnetic phase transition: using antiferromagnetic membranes FePS\textsubscript{3} as an example}

\author{Xiang Zhang}\email{zhang040321@gmail.com}
\affiliation{Beijing Academy of Quantum Information Science, Beijing, China}
\affiliation{Kavli Institute of Nanoscience, Delft University of Technology, Delft, The Netherlands}

\author{Makars Siskins}
\affiliation{Institute for Functional Intelligent Materials, National University of Singapore,\\4 Science Drive 2, Singapore 117544, Singapore}

\author{Yaroslav M. Blanter}
\affiliation{Kavli Institute of Nanoscience, Delft University of Technology, Delft, The Netherlands}

\date{\today}

\begin{abstract}
We developed a theoretical scheme of incorporating the magnetoelastic contribution into the thermal elastic dynamics for the thin membranes of 2D antiferromagnetic material with restricted geometry. We extended the elastic Gr\"uneisen relation into an effective version which includes the magnetic counterpart to the volume change of internal energy. Based on the specific heat and thermal conductivity from the elastic and magnetic origins we predicted the dependency of observables, such as effective Gr\"uneisen parameter, thermal expansion coefficient, and the damping factor, with respect to a wide range of temperature across the phase transition. Our model of analysis as been validated by applying to the case of FePS\textsubscript{3} flake resonator and the theoretical predictions fits well with the reported experiment data.
\end{abstract}

\maketitle

\section{Introduction}

In recent decades the 2D magnetic (van der Waals) layered materials have consistently attained the focus of research from both theoretical and experimental aspects~\cite{Cheng2019,Gong2017}. Compared to the three-dimensional counterpart, the 2D magnetic membranes constitute ideal platform to explore fundamental physics of magnetism and also its coupling to other degrees of freedom in the low dimensional regime~\cite{Hui2019}. The heterostructures build upon the 2D magnetism show susceptibility with respect to external stimuli leading to the emergent interfacial phenomena and novel spintronic devices~\cite{Cheng2019,Boona2014}. Within these materials, the FePS\textsubscript{3} compound is of particular interest because it is measured to be a 2D Ising model with zigzag antiferromagnetic (AFM) order in which the magnetic Fe atom constitute honeycomb lattice~\cite{Kargar2020,Lee2016}. Although the magnetic and electronic structure of this material has been studied intensively, there is limited understanding of its thermal properties and especially the magnetic contribution to the specific heat and thermal flux in the restricted geometry such as the thin membranes of several nanometers in thickness and micrometers in the planar dimension~\cite{Lancon2016,Wildes2012,Kargar2020}. The knowledge of its thermal properties is important for further application in spin-caloritronics~\cite{Boona2014} and also stands for another tool of investigating the magnetic phase transition apart from the Raman spectroscopy~\cite{Makars2020,Kargar2020}.

\begin{figure}[htb]
    \centering
    \includegraphics[width=\columnwidth]{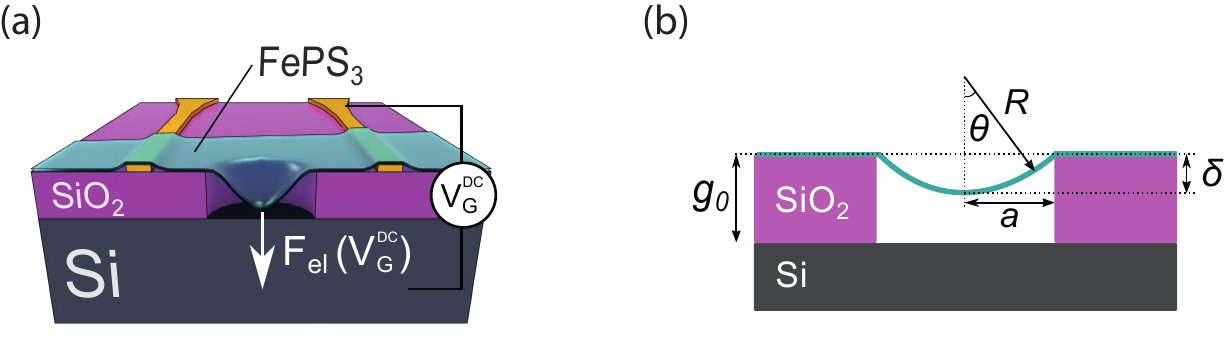}
    \caption[Schematic figure for the setup of FePS\textsubscript{3} resonator suspended on silicon substrate]{(a) Schematic figure for the FePS\textsubscript{3} resonator setup. The device is settled in nearly vacuum environment so that the thermal transfer through air damping can be ignored. Thermal expansion coefficient of the SiO\textsubscript{2} substrate is tiny and the silicon base is also small compared to the FePS\textsubscript{3}. The flake thickness is $h=45\,\mathrm{nm}$ and diameter $d=10\,\mu\mathrm{m}$. (b) Fixed gate voltage pushes down the membranes and as temperature increases the flake expands leading to a decrease of planar tension. Figure quoted from publication~\cite{Makars2020}.}
    \label{fig:makars_setup}
\end{figure}

In this Chapter, we extend the analysis of magnetoelastic coupling into a wide range of temperature beyond the phase transition, aiming at providing a theoretical explanation for the observed anomaly~\cite{Makars2020} in thermal transport of FePS\textsubscript{3} flake resonator. Showing in the Fig.~\ref{fig:makars_setup}, the membranes suspended over a cavity undergo a drum-like vibration whose eigenfrequency is related to the planar strain which can be tuned by the gate voltage and also by the environment temperature due to the thermal expansion. At a fixed gate voltage the membrane is pushed down, and the increase of temperature leads to the drop of strain that at around the Neel temperature $(T_N\approx 114\,\mathrm{K})$ the breaking of magnetic stiffness would soften this material and a sudden drop of resonance frequency has been observed~\cite{Makars2020}. Moreover, the vanishing of magnons as additional thermal carrier after $T>T_N$ would lead to a drop of the overall thermal conductivity which has been measured through the damping factor $Q^{-1}$ as function of temperature. 

In order to quantitatively explain experimental findings for thermal phenomena of the hybrid system, we develop a scheme of merging the magnetic contribution into the thermoelastic dynamics and predict the temperature dependence for observables including heat capacity, linear expansion coefficient, and damping factor for the clamped FePS\textsubscript{3} membranes. Starting from the non-magnetic thermoelastic free energy we firstly derive the expression for the damping factor $Q^{-1}$ of thin membrane/plate which turns out to be a function of the overall thermal expansion coefficient, the specific heat, and the thermal conductivity (See section~\ref{sec:bending_plate}). Then we derive the total specific heat $C_V$ which has origins including the phonon and magnon excitations and also the part of energy required to break the Ising coherence around phase transition. We calculate the thermal conductivity $\kappa$ as a sum of the phonon and magnon both as heat carriers and showed its magnitude are much smaller than the bulk compound because the limited particle lifetime due to the restricted geometry. Most importantly, by including the magnetoelastic Hamiltonian into the thermoelastic free energy we prove the total thermal expansion coefficient $\Tilde{\alpha}$ retains the usual formalism of Gr\"uneisen relation but with the incorporated effective Gr\"uneisen parameter $\Tilde{\gamma}$. It essentially describes the variation of internal energy including all the components ascribed to the volume change (See section~\ref{sec:hybrid}). Using real material parameters we fitted experimental measurements with our model of analysis. Good agreement with recent experiment data~\cite{Makars2020,takano2004} supports the validity of our results (See section~\ref{sec:model_validation}). The strong magnetic \textit{weight} as part of the internal energy for this geometry restricted membranes making it an ideal platform to study the optomechanics integrated with the magnetism tuning. It is also expected that the model developed in this work can be useful for further analysis in the 2D spin-caloritronic devices.

\section{Bending of thin plate with the temperature gradient}
\label{sec:bending_plate}

In order to calculate the damping coefficient $Q^{-1}$, we firstly have to solve the coupled dynamics including the degree of freedom from elasticity, magnetism, and temperature field. In the following section~\ref{sec:magnetoelastic}, one shall see that the contribution of magnetoelastic coupling can be incorporated into the effective thermoelastic coupling and the governing equations of motion can be narrowed to including only the dynamics of elastic vibration and temperature gradient. In this section, we deal with the round plate with its undeformed surface lying on the $X-Y$ plane and study its out-of-plane ($\hat{z}$) vibration. We use the cylinder coordinate $(r,\varphi,z)$ and assume its thickness $h$ is much smaller than the plate diameter $d$, i.e. $h\ll d$. The displacement $u_z$ and deformation $\epsilon_{ij}$ for plate are also considered to be small such that $u_i\ll h$ and $\epsilon_{ij}\ll h$. 

The displacement fields along $(\hat{r},\hat{\varphi})$ direction are respectively represented by $u_r$ meaning the radial extension and $u_\varphi$ meaning the circumferential distortion. One should note that $u_\varphi$ represents the displaced distance along the $\hat{\varphi}$ direction not the $\varphi$ itself, $u_\varphi=rd\varphi$. The strain tensor in cylinder coordinate is expressed in the form~\cite{Landau1986}
\begin{equation}
    \begin{split}
        &\epsilon_{rr}=\frac{\partial u_r}{\partial r},\;\epsilon_{\varphi\varphi}=\frac{1}{r}\frac{\partial u_\varphi}{\partial\varphi}+\frac{u_r}{r},\;\epsilon_{zz}=\frac{\partial u_z}{\partial z},\\
        &\epsilon_{\varphi z}=\frac{1}{2}\left(\frac{1}{r}\frac{\partial u_z}{\partial\varphi}+\frac{\partial u_\varphi}{\partial z}\right),\;\epsilon_{rz}=\frac{1}{2}\left(\frac{\partial u_r}{\partial z}+\frac{\partial u_z}{\partial r}\right),\\
        &\epsilon_{r\varphi}=\frac{1}{2}\left(\frac{\partial u_\varphi}{\partial r}-\frac{u_\varphi}{r}+\frac{1}{r}\frac{\partial u_r}{\partial \varphi}\right).
    \end{split}
\end{equation}
It is easy to show that according to the coordinate transformation, the relation $\epsilon_{rr}+\epsilon_{\varphi\varphi}+\epsilon_{zz}=\epsilon_{xx}+\epsilon_{yy}+\epsilon_{zz}$ holds, meaning the volume change, as it should, does not depends on the choice of coordination. Beyond this, the thermoelastic free energy~\cite{Landau1986}
\begin{equation}
    \small
    \begin{aligned}
        F(T)&=F_0(T)+\frac{1}{2}K_T\left(\epsilon_i^i\right)^2+\mu\sum_{ij}\left(\epsilon_{ij}-\frac{1}{3}\epsilon_i^i\,\delta_{ij}\right)^2\\
        &-K_T\alpha\left(T-T_0\right)\epsilon_i^i,
    \end{aligned}
    \label{eqn:thermoelastic_Free}
\end{equation}
and elastic tensor relation for isotropic material 
\begin{equation}
    \sigma_{ij}=K_T\epsilon_i^i\delta_{ij}+2\mu(\epsilon_{ij}-\frac{1}{3}\epsilon_i^i\delta_{ij}),\quad\epsilon_i^i=\sum_i\epsilon_{ii},
\end{equation}
also hold true in formalism for any orthogonal coordinates~\cite{Landau1986}. 

In order to effective describe the characteristic deformation of the 3D elastic body we establish a concept of neutral surface. Regarding to the bending of thin plate, one side is compressed (the concave side) while the opposite is extended (convex side). Between these two sides, there is a surface which has neither extension nor compression, i.e. $\epsilon_i^i=0$, and is referred as the \textit{neutral surface}. Mount the undeformed neutral surface onto the $z=0$ plane and based on the small deformation assumption, the displacement on the neutral surface is $u^0_r=0,\,u^0_\varphi=0,\,u^0_z=\zeta(r,\varphi,t)$ with $\zeta\ll h$. Due to the small deformation, the internal stress on $z$-th surface should be much smaller than the stress along the longitudinal direction, $\sigma_{iz}=0$, which leads to the hypotheses inside the bulk volume~\cite{Huang2005,Sun2008}
\begin{equation}
    \epsilon_{rz}=0,\quad \epsilon_{\varphi z}=0,\quad \sigma_{zz}=0.
\end{equation}
With the assumed neutral surface hypotheses, the displacement inside the plate can be expressed by the function of $\zeta$ that
\begin{equation}
    u_r=-z\frac{\partial \zeta}{\partial r},\quad u_\varphi=-\frac{z}{r}\frac{\partial\zeta}{\partial\varphi},\quad u_z=\zeta.
\end{equation}
and the remaining strain components are given by
\begin{equation}
    \small
    \begin{aligned}
        &\epsilon_{rr}=-z\frac{\partial^2 \zeta}{\partial r^2},\\
        &\epsilon_{r\varphi}=-z\frac{\partial}{\partial r}\left(\frac{1}{r}\frac{\partial \zeta}{\partial\varphi}\right),
    \end{aligned}
    \quad
    \begin{aligned}
        &\epsilon_{\varphi\varphi}=-z\left(\frac{1}{r}\frac{\partial \zeta}{\partial r}+\frac{1}{r^2}\frac{\partial^2 \zeta}{\partial\varphi^2}\right),\\
        &\epsilon_{zz}=\frac{z\sigma}{1-\sigma}\left(\frac{\partial^2 \zeta}{\partial r^2}+\frac{1}{r}\frac{\partial \zeta}{\partial r}+\frac{1}{r^2}\frac{\partial^2 \zeta}{\partial \varphi^2}\right).
    \end{aligned}
\end{equation}
Define the Laplace operator on the plane 
\begin{equation}
    \Delta=\frac{\partial^2}{\partial r^2}+\frac{1}{r}\frac{\partial}{\partial r}+\frac{1}{r^2}\frac{\partial^2}{\partial \varphi^2},
\end{equation}
then $\epsilon_{rr}+\epsilon_{\varphi\varphi}=-z\Delta\zeta$ and $\epsilon_{zz}=\frac{z\sigma}{1-\sigma}\Delta\zeta$. For the case of axial symmetric plate it is reasonable to assume $\zeta=\zeta(r,t)$ which does not depends on the polar angle $\varphi$, then the strain can be even simplified into
\begin{equation}
    \begin{aligned}
        &\epsilon_{rr}=-z\frac{\partial^2 \zeta}{\partial r^2},\quad\epsilon_{\varphi\varphi}=-\frac{z}{r}\frac{\partial\zeta}{\partial r},\\
        &\epsilon_{zz}=\frac{z\sigma}{1-\sigma}\Delta\zeta,\quad\Delta=\frac{\partial^2}{\partial    r^2}+\frac{1}{r}\frac{\partial}{\partial r},
    \end{aligned}
    \label{eqn:2Dplate_strain}
\end{equation}
and other components equals to zero. Substituting the strain tensor into the thermoelastic free energy (Eq.~\ref{eqn:thermoelastic_Free}) one can derive its expression as the function of $\zeta$
\begin{widetext}
\begin{equation}
    F_{\mathrm{el}}=\int_0^{2\pi}d\varphi\int_0^R rdr\frac{Yh^3}{12(1-\sigma^2)}\left[(1+\sigma)\frac{\alpha}{3}\,I_T\Delta\zeta+\frac{1}{2}\left({\zeta''}^2+\frac{1}{r^2}{\zeta'}^2+\frac{2\sigma}{r}\zeta'\zeta''\right)\right],
\end{equation}
\end{widetext}
where the thermal inertia
\begin{equation}
    I_T(r)=\frac{12}{h^3}\int_{-h/2}^{h/2}z\,\theta(r,z)\,dz,
    \label{eqn:Inertia_temp}
\end{equation}
in which $\theta=T-T_0$ is the small differences between the temperature within the plate $T$ and the environment temperature $T_0$. The internal force exerted on to the volume element of unit surface is $f_\zeta=-\delta F_{\mathrm{el}}\big/\delta\zeta$ and the equation of motion for the vibration of the circular plate is
\begin{equation}
    \rho h\frac{\partial^2\zeta}{\partial t^2}+\frac{Yh^3}{12(1-\sigma^2)}\left[\Delta\Delta\zeta+(1+\sigma)\alpha/3\,\Delta I_T\right]=0.
    \label{eqn:2Dplate_elastic}
\end{equation}

As for the dynamics of temperature field the heat diffusion equation is a rephrase of energy conservation, that is the heat absorption equals to the energy flows $T\frac{\partial S}{\partial t}=-\nabla\cdot\boldsymbol{q}=\kappa\Delta T$ with $\boldsymbol{q}=-\kappa\nabla T$ is the thermal flux and $\kappa$ is the heat conduction coefficient~\cite{McPherson2019}. From the thermoelastic coupling we understand that the heat absorption leads to not only increase of particle motion but also the volume expansion, $dS=dS_0(T)+K_T\alpha\epsilon_i^i$. Applying the relation $\partial S_0\big/\partial T=\rho C_V\big/T$, we have $\rho C_V\partial T\big/\partial t=\kappa\Delta T-K_T\alpha T_0\partial\epsilon_i^i\big/\partial t.$ The equation of motion for describing the dynamics of small temperature differences within the plate has the general form
\begin{equation}
    \kappa\Delta\theta+\kappa\frac{\partial^2\theta}{\partial z^2}=\rho C_V\frac{\partial\theta}{\partial t}+K\alpha T_0\frac{\partial\epsilon_i^i}{\partial t}.
\end{equation}
As from Ref.~\cite{Lifshitz2000} we make an approximation that the temperature gradient is small in the longitudinal direction compared to the vertical direction, $\Delta\theta\ll \partial^2\theta/\partial z^2$. Combing the strain components from Eq.~\ref{eqn:2Dplate_strain} the governing equation for the dynamics of temperature field in thin plate then becomes
\begin{equation}
    \kappa\frac{\partial^2\theta}{\partial z^2}=\rho C_V\frac{\partial\theta}{\partial t}-zK_T\alpha T_0\frac{1-2\sigma}{1-\sigma}\frac{\partial\Delta\zeta}{\partial t}.
    \label{eqn:2Dplate_thermal}
\end{equation}

Inserting the ansatz solution $\zeta=\zeta_0e^{i\omega t}$ and $\theta=\theta_0e^{i\omega t}$ into the Eq.~\ref{eqn:2Dplate_thermal} we have the equation for temperature field which can be solved by the boundary condition that there is no thermal conduction on the top and bottom surface,
\begin{equation}
    \frac{\partial\theta_0}{
    \partial z}=0\quad\mathrm{at}\;z=\pm\frac{h}{2}.
\end{equation}
The solved temperature profile across the plate is given by
\begin{equation}
    \theta_0(r,z)=\frac{K_T\alpha T_0}{\rho C_V}\frac{1-2\sigma}{1-\sigma}\left[z-\frac{\sin{(mz)}}{m\cos(mh/2)}\right]\Delta\zeta_0,
\end{equation}
with the wave vector
\begin{equation}
    m=\sqrt{-\frac{i\omega\rho C_V}{\kappa}}=(1-i)\sqrt{\frac{\omega\rho C_V}{2\kappa}}.
\end{equation}
Applying this temperature profile into the moment of inertia (Eq.~\ref{eqn:Inertia_temp}) and the elastic equation of motion (Eq.~\ref{eqn:2Dplate_elastic}) becomes an eigen-equation
\begin{equation}
    \begin{aligned}
        \rho h\omega^2 \zeta_0&=\frac{Yh^3}{12(1-\sigma^2)}[1+\Delta_Y(1+f(\omega))]\Delta\Delta \zeta_0\\
        &=\frac{Y_\omega h^3}{12(1-\sigma^2)}\Delta\Delta \zeta_0,
    \end{aligned}
    \label{eqn:eigen_w0}
\end{equation}
with the modified Young's modulus $Y_\omega=[1+\Delta_Y(1+f(\omega))]$ is frequency-dependent and the adiabatic degree $f(\omega)$ ranges from $-1$ to $0$ for low and high vibrating frequency identifying the isothermal and adiabatic extremes,
\begin{equation}
    f(\omega)=\frac{24}{m^3h^3}\left[\frac{mh}{2}-\tan{\left(\frac{mh}{2}\right)}\right].
\end{equation}
The quantity $\Delta_Y$ which is a measure of \textit{thermal relaxation strength} acquires the from
\begin{equation}
    \Delta_Y=\frac{1+\sigma}{1-\sigma}\frac{Y\alpha^2T_0}{\rho C_V}.
\end{equation}
Letting $12(1-\sigma^2)\rho\omega^2\big/Y_\omega h^2=q^4$, then the Eq.~\ref{eqn:eigen_w0} becomes $\Delta\Delta \zeta_0=q^4\zeta_0$ and can be solved by $\Delta \zeta_0=q^2\zeta_0$ with $\zeta_0=AJ_0(qr)+BY_0(qr)+CI_0(qr)+DK_0(qr)$. Here $(J_0,Y_0,I_0,K_0)$ are the first and second Bessel functions of the zero-th order respectively. Due to the finite value of $\zeta_0$ at $r=0$, the $B=D=0$ and $\zeta_0=AJ_0(qr)+CI_0(qr)$ with the coefficient $(A,C)$ to be defined by the boundary condition. For the case of clamped plate, the boundary condition ($a$ is plate radius) has the form $\zeta_0\big|_{r=a}=0,\;\partial\zeta_0\big/\partial r\big|_{r=a}=0$, which can be satisfied by $(q_na)^2\equiv \mathcal{C}_n=\{10.21,39.38,89.10,\cdots\}$. The complex eigenfrequency then reads
\begin{equation}
    \omega=\omega_0\sqrt{1+\Delta_Y(1+f(\omega_0))},
\end{equation}
with the unperturbed eigenfrequency for the $n$-th vibration mode is
\begin{equation}
    \omega_0=q_n^2h\sqrt{\frac{Y}{12\rho(1-\sigma^2)}}=\mathcal{C}_n\frac{h}{a^2}\sqrt{\frac{Y}{12\rho(1-\sigma^2)}}.
    \label{eqn:frequency_plate}
\end{equation}

Due to the complex value of frequency $\omega$, the time dependency $e^{i\omega t}$ of physical quantity decays along with the oscillation. Assuming $\omega=\omega_0(1+i\eta)$ then the displacement decays as $\zeta(t)\sim e^{i\omega_0t}e^{-\eta\omega_0t}$. The damping for this oscillating system is captured by the damping factor $Q^{-1}$ which is defined to be the ratio of energy loss per radian to the energy stored in the oscillator. Because the oscillating energy is quadratic to the displacement field so we have $E(t)\sim e^{-2\eta\omega_0t}$ leading to the fractional energy loss per radian is $1-e^{-2\eta}\approx 2\eta$. Thus the system damping for elastic oscillator is qualified by the $Q^{-1}=2|\mathrm{Im}(\omega)\big/\mathrm{Re}(\omega)|$. Shortening the parameter $mh$ within the function $f(\omega)$ into a single variable $\xi$~\cite{Lifshitz2000}
\begin{equation}
    \xi=h\sqrt{\frac{\omega_0\rho C_V}{2\kappa}},
\end{equation}
the thermoelastic damping $Q^{-1}$ can be derived as
\begin{widetext}
    \begin{equation}
        Q^{-1}=\Delta_Y\left(\frac{6}{\xi^{2}}-\frac{6}{\xi^{3}}\frac{\sinh\xi+\sin\xi}{\cosh\xi+\cos\xi}\right)=\frac{1+\sigma}{1-\sigma}\frac{Y\alpha^2 T_0}{\rho C_V}\left(\frac{6}{\xi^{2}}-\frac{6}{\xi^{3}}\frac{\sinh\xi+\sin\xi}{\cosh\xi+\cos\xi}\right).
    \label{eqn:Qinverse_plate_nomag}
    \end{equation}
\end{widetext}
Since the thermoelastic variables such as $\alpha$, $\kappa$ and $C_V$ are temperature dependent, it is easy to understand the damping factor $Q^{-1}$ also changes with $T_0$ and it will show anomaly in the present of second order phase transition with which the specific heat $C_V$ has observed discontinuity. For convenience, in the following we will replace the environment temperature $T_0$ by the symbol $T$ with consensus.

\section{Thermal observables for elastic plate hybrid with magnetic phase transition}
\label{sec:hybrid}

In this section we study the thermal observables for the elastic plate hybrid with magnetism for a wild range of temperature across the phase transition. To this aim we start with deriving the heat capacity and thermal conductivity due to the bosons. Then we shows the incorporation of magnetoelastic coupling into the effective thermoelastic free energy and derive the effective expansion coefficient $\Tilde{\alpha}$ and damping factor $Q^{-1}$ for the thermal-magnetic-elastic vibrating system.

In general, below the phase transition the material's heat capacity $C=dQ\big/dT$ comes from the thermal excitation of the bosons, which are quasi-particles mainly the phonons for ordinary insulators and also include magnons for FM and AFM materials. If the temperature is homogeneous then the Bose-Einstein density of excited bosons is uniformly distributed across the material. However, the existence of temperature field leads to the excess number of quasi-particles staying out of equilibrium and then transport according to the temperature gradient. If the environment temperature is close to the range of magnetic phase transition, the coherence of precession between the neighbouring spins breaks down and an additional contribution to the specific heat should be taken into account. The decaying of magnetization $\boldsymbol{M}$ as the heating procedure leads to an accompanying decrease of the effective exchange field $H_E$ and anisotropy field $H_A$ in magnon's dispersion equation
\begin{equation}
    \omega_{\boldsymbol{k}}=\gamma\mu_0\sqrt{H_A^2+2H_EH_A+H_E^2(1-\psi_{\boldsymbol{k}}^2)},
    \label{eqn:AFSW}
\end{equation}
in which $\psi_{\boldsymbol{k}}$ is the structure factor defined by $\psi_{\boldsymbol{k}}=(1/z)\sum_{\boldsymbol{\delta}} e^{i\boldsymbol{k}\cdot\boldsymbol{\delta}}$ and $\boldsymbol{\delta}$ is the vector connecting the $z$ nearest neighbouring spins of opposite orientations. This energy renormalization~\cite{Rezende2014,Shen2018} should also be incorporated into the calculation of magnon's specific heat and thermal conductivity.

Mathematically the heat capacity due to the bosons is
\begin{equation}
    C_V=\frac{1}{V}\frac{\partial}{\partial T}\sum_{\boldsymbol{k}}\hbar\omega_{\boldsymbol{k}}\Bar{n}_{\boldsymbol{k}},\quad \Bar{n}_{\boldsymbol{k}}=\frac{1}{e^{\beta\hbar\omega_{\boldsymbol{k}}}-1},
    \label{eqn:general_capacity}
\end{equation}
where $\Bar{n}_{\boldsymbol{k}}$ is the Bose-Einstein's equilibrium amount of bosons of energy $\hbar\omega_{\boldsymbol{k}}$. The thermal conductivity is defined as the coefficient for heat flux due to the temperature gradient, $\boldsymbol{q}=-\kappa\nabla T$. From kinetic transfer theory this thermal flux can be calculated by
\begin{equation}
    \begin{aligned}
        \boldsymbol{q}&=-\frac{1}{V}\sum_{\boldsymbol{k}}\hbar\omega_k\boldsymbol{v}_k(\tau_k\boldsymbol{v}_k\cdot\nabla\Bar{n}_k)\\
        &=-\frac{1}{V}\frac{\partial}{\partial T}\sum_{\boldsymbol{k}}\hbar\omega_k\Bar{n}_k\tau_k(\nabla T\cdot\boldsymbol{v}_k)\boldsymbol{v}_k,
    \end{aligned}
    \label{eqn:general_flux}
\end{equation}
in which an isotropic $\kappa$ can be extracted if the particle velocity $\boldsymbol{v}_k$ is homogeneous to each direction. However, if the particle velocity has directional bias then $\kappa$ depends on the orientation and thermal transfer shows anisotropy. In the simplest case, if the particle's lifetime $\tau_k=\tau_0$ and velocity $\boldsymbol{v}_k=\Bar{v}$ does not depends on wavevector $\boldsymbol{k}$ we see that the thermal flux can be simplified as $\boldsymbol{q}=-\frac{\Bar{v}^2\tau_0}{V}\frac{\partial}{\partial T}\sum_{\boldsymbol{k}}\hbar\omega_k\Bar{n}_k\nabla T\equiv -C\Bar{v}^2\tau_0\cdot\nabla T$ leading to the simple form $\kappa=C\Bar{v}^2\tau_0$.
Once we know the dispersion relation $\omega_k$ and the lifetime $\tau_k$ for the mobile quasi-particles we can determine specific heat $C_V$ and thermal conductivity $\kappa$ at least in a numerical way. 

The elastic specific heat and thermal conductivity can be derived from the statistics of low lying phonon modes (acoustic modes) based on the general equations of \ref{eqn:general_capacity} and \ref{eqn:general_flux} with the sound wave dispersion relation $\omega_{\boldsymbol{k}}=\Bar{v}\boldsymbol{k}$, and $\Bar{v}$ is the Debye averaged acoustic velocity
\begin{align}
    \begin{split}
        &C_\mathrm{db}(T)=\frac{\hbar^2}{2\pi}\frac{3}{k_BT^2}\int_0^{k_\mathrm{db}}dk\frac{k\omega_k^2e^{\beta\hbar\omega_k}}{(e^{\beta\hbar\omega_k}-1)^2},\\ &\kappa_\mathrm{db}(T)=\frac{\hbar^2}{4\pi}\frac{3\Bar{v}^2}{k_BT^2}\int_0^{k_\mathrm{db}}dk\frac{\tau_k k\omega_k^2e^{\beta\hbar\omega_k}}{(e^{\beta\hbar\omega_k}-1)^2}.
    \end{split}
    \label{eqn:cph_2D}
\end{align}
The elastic zone boundary can be defined by the Debye temperature as $k_\mathrm{db}=k_BT_\mathrm{db}\big/\hbar\Bar{v}$. Note that here we have assumed the elastic lattice is of 2D while the vibration is still 3 dimensional. It can be adjusted easily to the longitudinal or shear polarization only by replacing the factor $3$ into $1$ or $2$ respectively.

\subsection{Specific heat and thermal conduction due to the magnon excitations}

Specific to the 2D AFM material, we set the external field $H_0=0$ and according to Eq.~\ref{eqn:AFSW} the dispersion relation depends on the direction of wave vector $\boldsymbol{k}$. Contrary to detailed treatment as in the Refs.~\cite{Lancon2016}, in this work we simplify the detailed 2D lattice structure and make the homogeneous assumption ($a=b$) such that we rephrase the $\psi_{\boldsymbol{k}}$ of Eq.~\ref{eqn:AFSW} into $\psi_{k}=\cos{(\pi k\big/2k_m)}$ being isotropic with $k\in[0,k_m]$ is limited to the first Brillouin zone and $k_m$ is defined from the spherical energy boundary assumption~\cite{Simon2016,Rezende2014},
\begin{equation}
    \sum_{\boldsymbol{k}}=\frac{V}{(2\pi)^2}\int d\boldsymbol{k}=\frac{Na^2}{(2\pi)^2}\int_0^{k_m} 2\pi k dk=N,
\end{equation}
such that $k_ma/2=\sqrt{\pi}$.
Thus, the dispersion relation for AFM magnon becomes
\begin{equation}
    \hbar\omega_k=\gamma\mu_0 H_E\sqrt{\sin^2{(\pi k/2k_m)}+\eta^2+2\eta},
    \label{eqn:2Dmagnon}
\end{equation}
with $\eta=H_A\big/H_E$ is the ratio of anisotropy field to the exchange field. For some AFM material such as the RbMnF\textsubscript{3} which has very small anisotropy $H_A=4.5\;\mathrm{Oe}$, the exchange field is as large as $H_E=830\;\mathrm{kOe}$ leading to $\eta\approx 0$ and it is considered as a typical 3D Heisenberg antiferromagnet~\cite{Ortiz2014,Cole1966}. For other materials such as the FeF and the FePS\textsubscript{3} used in our experiment the magnetic anisotropy is strong and comparable to the exchange field, resulting in $\eta\gtrapprox 1$ which makes them a quasi Ising system~\cite{Ortiz2014,Wang2016}.

As environment temperature goes up, the spontaneous magnetization $M(T)$ decays because the thermal magnon excitation~\cite{Rezende2014,Rezende2019} and also the decoherence between neighbouring spins for the $T\lessapprox T_N$. Since $M=-g\mu_BNS$, the effective spin magnitude $S(T)$ decays which results in the decreasing of $H_E=2Sz|J|\big/\mu_0\gamma$ and $H_A=2SA\big/\mu_0\gamma$ in the dispersion relation. As a consequence, the temperature dependence for the $\omega(T)$ should be taken into account in deriving the magnon specific heat and thermal conductivity. For simple treatment one can apply the molecular field approximation (mean field theory) in which the magnetization $M(T)=M_0B(x)$ with $B(x)$ is the Brillouin function and $x=\mu_0n_wM(T)g\mu_BS_0\big/k_BT$ is the normalized energy~\cite{Coey2010}. Although this mean field approach does not provide the correct magnetization around phase transition, it leads to good results of magnon spectra at temperatures $T<0.8T_N$~\cite{Rezende2014,Shen2018}. 

\begin{figure}
    \centering
    \includegraphics[width=\columnwidth]{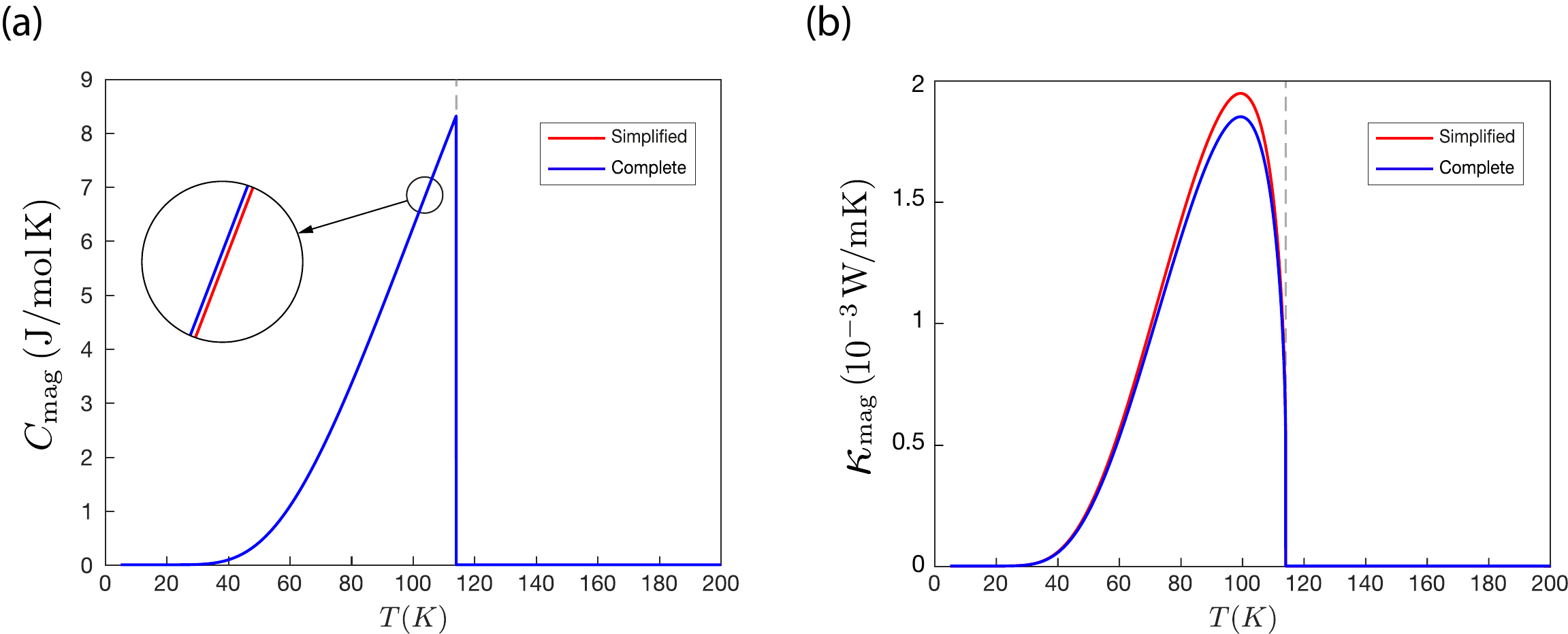}
    \caption[Comparison for the magnon's specific heat and thermal conductivity derived from complete and simplified integral.]{(a) The magnon's specific heat and (b) thermal conductivity derived from the complete 2-D integral and the simplified 1-D integral respectively. Here we assumed the lifetime for magnon is approximately $1.8\,\mathrm{ps}$ and does not depends on the modes for simplicity. The results indicate the difference between these two integral strategy is small and we can use the simplified version for further calculations.}
    \label{fig:integral_compare}
\end{figure}

With the derived dispersion relation, the heat capacity due to thermal magnons excitation in the 2D AFM model is
\begin{align}
    \begin{split}
        C_\mathrm{mag}(T)&=\frac{\hbar^2}{2\pi}\frac{1}{k_BT^2}\int_0^{k_m}dk \frac{k\,\omega^2_k\cdot e^{\beta\hbar\omega_k}}{(e^{\beta\hbar\omega_k}-1)^2}\\
        &=\frac{\hbar^2k_m^2}{2\pi k_B}\frac{1}{T^2}\int_0^1dq \frac{q\,\omega^2_q\cdot e^{\beta\hbar\omega_q}}{(e^{\beta\hbar\omega_q}-1)^2},
    \end{split}
    \label{eqn:cm_2D}
\end{align}
where the explicit temperature dependence on $\omega_k(T)$ has been suppressed and we replace $k$ with the normalized wave vector $q=k\big/k_m$ ranging from $0$ to $1$. As a comparison, we plot the $C_\mathrm{mag}(T)$ derived from the 2-D integral of $(q_x,\,q_y)$ of the dispersion relation $\hbar\omega_{\boldsymbol{q}}=\gamma\mu_0 H_E\sqrt{(1-\psi_{\boldsymbol{q}}^2)+\eta^2+2\eta}$ with $\psi_{\boldsymbol{q}}=\cos{(q_x\pi/2)}\cos{(q_y\pi/2)}$. Showing in Fig.~\ref{fig:integral_compare}(a) we see the complete 2-D integral and the simplified one result in almost exactly the same curve which validates that we can indeed ignore the direction of $(q_x,\,q_y)$ and shorten the $\psi_{\boldsymbol{q}}$ into 1-D integral on $q$ with the $\psi_q=\cos{(q\pi/2)}$.

For the magnon's thermal conductivity, it is defined from the heat flux (Eq.~\ref{eqn:general_flux}) that 
\begin{equation}
    \boldsymbol{q}=\frac{-1}{(2\pi)^2}\int\,d\boldsymbol{k}\frac{1}{k_BT^2}\frac{\tau_k\,(\hbar\omega_k)^2\,e^{\beta\hbar\omega_k}}{(e^{\beta\hbar\omega_k}-1)^2}(\nabla T\cdot\boldsymbol{v}_k)\boldsymbol{v}_k.
\end{equation}
Using the 2-D dispersion relation we derive the velocity of magnons to be
\begin{equation}
    \begin{aligned}
        (v_x,v_y)&=\frac{\gamma\mu_0 H_E}{2\hbar k_m}\frac{\pi\psi_k}{\sqrt{(1-\psi_k^2)+\eta^2+2\eta}}\times\\
        &\left(\sin{\frac{k_x\pi}{2k_m}}\cos{\frac{k_y\pi}{2k_m}},\;\cos{\frac{k_x\pi}{2k_m}}\sin{\frac{k_y\pi}{2k_m}}\right),
    \end{aligned}
\end{equation}
with which the integral can be performed as
\begin{equation}
    \int (\boldsymbol{\nabla T}\cdot\boldsymbol{v}_k)\boldsymbol{v}_k\,d\boldsymbol{k}=\boldsymbol{\nabla T}\,\frac{1}{2}\int(v_x^2+v_y^2)\,d\boldsymbol{k},
\end{equation}
where we have used the fact that $\int v_x^2d\boldsymbol{k}=\int v_y^2\boldsymbol{k}$ due to the symmetry consideration. From the heat flux expression we extract the thermal conductivity and written into the form of 2-D integral,
\begin{widetext}
    \begin{equation}
    \begin{aligned}
        \kappa_\mathrm{mag}&=\left(\frac{\gamma\mu_0 H_E}{8}\right)^2\frac{1}{k_BT^2}\int \frac{\tau_k\, e^{\beta\hbar\omega_q}}{(e^{\beta\hbar\omega_q}-1)^2}\left[\left(\frac{\gamma\mu_0 H_E}{\hbar}\right)^2-\left(\omega_q^2-\omega_0^2\right)\right](1-\cos{q_x\pi}\cdot\cos{q_y\pi})d\boldsymbol{q},\\
        &=\left(\frac{\gamma\mu_0 H_E}{8}\right)^2\frac{\pi}{k_BT^2}\int_0^1\,\frac{\tau_k\,\omega_q^2e^{\beta\hbar\omega_q}}{(e^{\beta\hbar\omega_q}-1)^2}\frac{q\sin^2{q\pi}}{\sin^2{q\pi/2}+\eta^2+2\eta}dq,
    \end{aligned}
    \label{eqn:km_2D}
    \end{equation}
\end{widetext}
where the second equality is reached by the homogeneous lattice assumption that the integral can be simplified into the 1-D and $\hbar\omega_q=\gamma\mu_0 H_E\sqrt{\sin^2{q\pi/2+\eta^2+2\eta}}$. Showing in Fig.~\ref{fig:integral_compare}(b) we notice again the difference between complete 2-D integral and the simplified one is small enough for the case of $\kappa_\mathrm{mag}$, and in the following we shall use the 1-D integral of $q$ for the thermal observables. At low temperatures the heat capacity and thermal conductivity share the same growing curve due to the fact that $\kappa\approx Cvl=Cv^2\tau$ and $v,\tau$ are almost constant for small $T$. At large enough temperatures the exchange fields $H_E(T)$ decays in step with the decreasing of $M(T)$ which leads to the softening of magnons and after phase transition there is no existence of magnons. Thus we see the drop of $C_\mathrm{mag}$ and $\kappa_\mathrm{mag}$ for $T>T_N$. Additionally the particle's lifetime (or its inverse $\tau^{-1}=\eta$ the relaxation rate) plays an important role in their transport properties. In general, the relaxation rate for various particles, either bosons or fermions, comes from several origins~\cite{Neelmani1972} that $\eta=\eta_\mathrm{bd}+\eta_\mathrm{pt}+\eta_\mathrm{nlsc}$, with $\eta_\mathrm{bd}$ is the boundary deflection by material edges, $\eta_\mathrm{pt}$ is the scattering with the point defects, and $\eta_\mathrm{nlsc}$ stands for the non-linear scattering among particles themselves. Usually $\eta_\mathrm{bd}+\eta_\mathrm{pt}=\eta_0\equiv\tau_0^{-1}$ is a constant which does not depend on wavevector $k$ and temperature $T$. The non-linear scattering has several origins for different particles but it is generally proportional to $T$ for the 3-particle scattering and $T^2$ for the 4-particle scattering process~\cite{Kittel2004,Rezende2014-2}. Therefore $\eta_k=\eta_0(1+b_kT+c_kT^2)$ and the coefficients can be calculated by studying the detailed process. However, in this work of membranes setup both the phonon and magnon's lifetime are limited by the defect and boundary scattering~\cite{Dolleman2017}. Therefore we shall ignore the non-linear scattering between quasi-particles and claim the lifetime $\tau=\tau_0$ is a constant which does not depends on the wave vector nor the temperature.

\subsection{Specific heat due to the break of spin coherence around phase transition}

\begin{figure}
    \centering
    \includegraphics[width=\columnwidth]{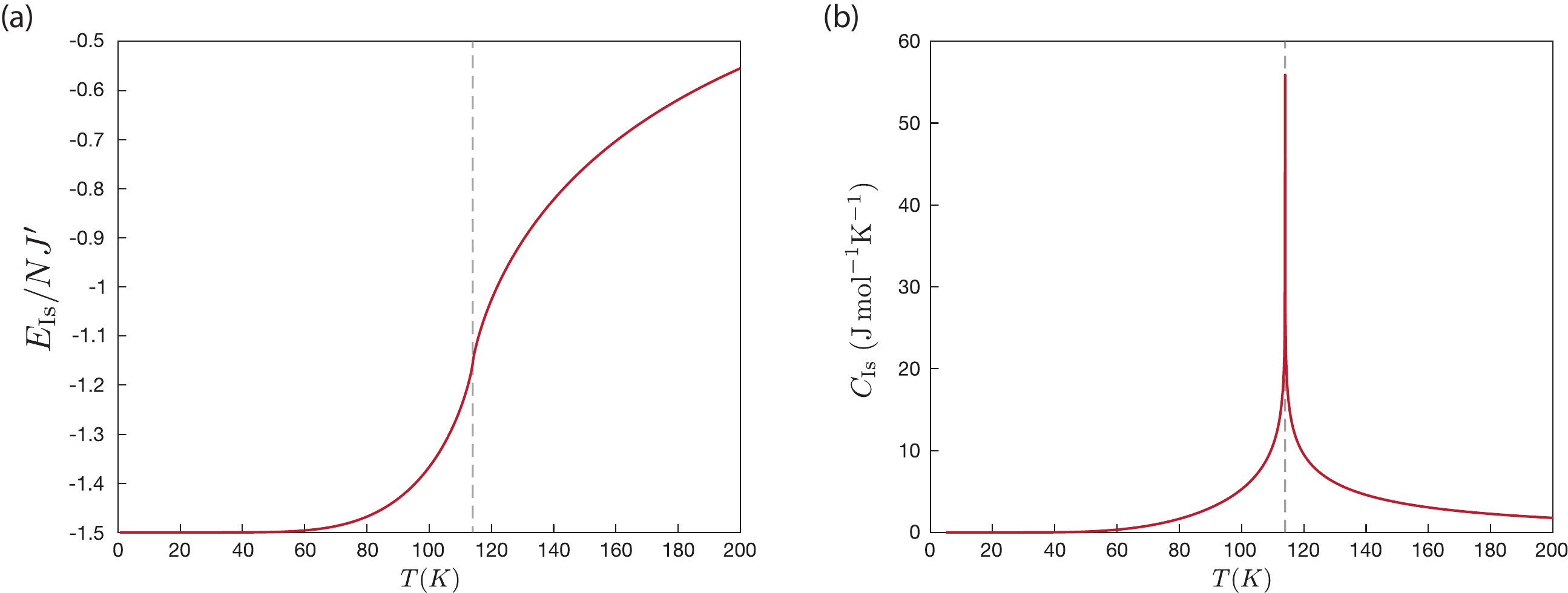}
    \caption[Magnetic specific heat for the 2D Ising model of antiferromagnetic honeycomb lattice.]{The temperature dependence of magnetic energy and specific for 2D Ising honeycomb lattice, numerically calculated according to the Eq.~\ref{eqn:cm_honey}. One can check that at $T=0$, $K\rightarrow\infty$ then $E_\mathrm{Is}\big/NJ'\rightarrow-3/2$ as expected because the coordination number for honeycomb lattice is $3$. }
    \label{fig:C_Ising}
\end{figure}

As the environment temperature close to the phase transition regime the magnetic specific heat is dominated by energy absorption for the breaking of spin coherence and due to the nature of second order phase transition the anomaly of $C_M$ near $T_N$ should be expected~\cite{Coey2010}. The derivation for anomaly of $C_M$ depends on the detailed lattice structure. In this chapter we focus on the material FePS\textsubscript{3} which is an Ising-type 2D antiferromagnet of the honeycomb (hexagon) lattice~\cite{Lee2016,Wildes2012,Jianmin2021,Lancon2016}. According to the references~\cite{Houtappel1950,Pathria2011,Matveev1996}, the partition function for honeycomb lattice reads
\begin{widetext}
    \begin{equation}
        \frac{1}{N}\log{Z(T)}=\log{2}+\frac{1}{16\pi^2}\int_0^{2\pi}\int_0^{2\pi}\,d\theta_1d\theta_2\log{\left[\cosh^3{2K}+1-\sinh^2{2K}\cdot P_{\boldsymbol{\theta}}\right]},
        \label{eqn:partition_honey}
    \end{equation}
\end{widetext}
where $K=J'\big/k_BT\equiv\beta J'$ is the normalized temperature in which $J'$ is the effective coupling energy from the exchange Hamiltonian $H=-2J\sum\boldsymbol{S}_i\cdot\boldsymbol{S}_j\equiv-J'\sum\hat{\boldsymbol{S}}_i\cdot\hat{\boldsymbol{S}}_j$, thus $J'=2JS^2$~\cite{Pathria2011}. The integrand parameter is $P_{\boldsymbol{\theta}}=\cos{\theta_1}+\cos{\theta_2}+\cos{(\theta_1+\theta_2)}$~\cite{Matveev1996}.
The critical point for honeycomb lattice is reached as $\sinh{2K_c}=\sqrt{3}$ and the Neel temperature is
\begin{equation}
    T_N=\frac{2J'}{k_B\log{(2+\sqrt{3})}}.
    \label{eqn:Neel}
\end{equation}
Thus one can derive the effective coupling energy $J'$ based on the measured Neel temperature. Following the procedures of differentiating $E_\mathrm{Is}=-\frac{d\log{Z}}{d\beta}$ and $C_\mathrm{Is}=\frac{dE_\mathrm{Is}}{dT}$, we have the specific heat due to the breaking of spin coherence reads
\begin{align}
    \begin{split}
        &\frac{1}{Nk_B}C_\mathrm{Is}(T)=\frac{K^2}{16\pi^2}\int_0^{2\pi}\int_0^{2\pi}d\theta_1d\theta_2\,\Bigg\{\\
        &\frac{6\sinh{4K}\sinh{2K}-4\cosh{4K}(2P_{\boldsymbol{\theta}}-3\cosh{2K})}{\cosh^3{2K}+1-\sinh^2{2K}\cdot P_{\boldsymbol{\theta}}}\\
        &-\frac{\sinh^2{4K}(2P_{\boldsymbol{\theta}}-3\cosh{2K})^2}{\left[\cosh^3{2K}+1-\sinh^2{2K}\cdot P_{\boldsymbol{\theta}}\right]^2}\Bigg\}.
    \end{split}
    \label{eqn:cm_honey}
\end{align}
The temperature dependence of the Ising type magnetic energy and specific heat is shown in Fig.~\ref{fig:C_Ising} Although there is no macroscopic magnetization above $T_N$, the short range spin correlations still persist and shall finally vanish to zero for large enough temperature. The total magnetic specific heat for our material FePS\textsubscript{3} is then a sum of the Ising and magnon's contribution $C_M=C_\mathrm{Is}+C_\mathrm{mag}$.

\subsection{Incorporating the magnetoelastic coupling into thermoelastic free energy}
\label{sec:magnetoelastic}

The previous analysis for the damping of thin plate in section~\ref{sec:bending_plate} does not involve the magnetic effect on the heat transportation. In nature, the magnetoelastic coupling should be included into the total free energy and the analysis of the hybrid of magnetic, elastic, and thermal dynamics should be considered~\cite{Prasai2017,Argyle1967}. According to Refs.~\cite{Shapira1978,Shapira1976}, the magnetoelastic coupling energy in general has the form
\begin{equation}
    F_{\mathrm{MEC}}=-Nz\frac{\partial J}{\partial V}\expval{\boldsymbol{S}_i\cdot\boldsymbol{S}_j}\epsilon_i^i,
    \label{eqn:F_MEC}
\end{equation}
in which $N$ is the number of spins per unit volume, $z$ is the coordination number, and $\epsilon_i^i$ is the fractional volume change. The two-spin correlation function $\expval{\boldsymbol{S}_i\cdot\boldsymbol{S}_j}$ indicates the average over space and time for any two nearest neighbouring spins. Incorporating this free energy into the thermal elastic one (Eq.~\ref{eqn:thermoelastic_Free}) we write down the total free energy taking both the thermal and magnetic elastic coupling into consideration
\begin{equation}
    \begin{aligned}
        F&=F_0-K_T\alpha_E\,\theta\,\epsilon_i^i+\frac{1}{2}K_T\left(\epsilon_i^i\right)^2+\mu\sum_{ij}\left(\epsilon_{ij}-\frac{1}{3}\epsilon_i^i\,\delta_{ij}\right)^2\\
        &-Nz\frac{\partial J}{\partial V}\expval{\boldsymbol{S}_i\cdot\boldsymbol{S}_j}\epsilon_i^i.
    \end{aligned}
\end{equation}
Note that we have replaced $\alpha$ in Eq.~\ref{eqn:thermoelastic_Free} with the symbol $\alpha_E$ in order to highlight its elastic origin. The strain in equilibrium can be derived from $\partial F\big/\partial\epsilon_i^i=0$, leading to the combined effect on volume changes due to both thermal expansion and magnetostriction
\begin{equation}
    \epsilon_i^i=\alpha_E\theta-\beta_TNz\frac{J}{V}\gamma_M\expval{\boldsymbol{S}_i\cdot\boldsymbol{S}_j},
    \label{eqn:thermal_expansion}
\end{equation}
where the magnetic Gr\"uneisen constant $\gamma_M$ describing the volume dependence on the exchange coupling strength has the form~\cite{Shapira1978,Gomes2019}
\begin{equation}
    \gamma_M=-\frac{V}{J}\frac{\partial J}{\partial V}.
\end{equation}
The part of volume changes due to the magnetostriction is proportional to the two-spin correlation function which can be changed by the variation of either temperature or external field. In this chapter we assume there is no external field applied onto the plate, then the temperature increase leads to the decaying of spin correlation and results in the magnetostriction expansion. Since the magnetic energy derived from the Heisenberg Hamiltonian $H=-2J\sum\boldsymbol{S}_i\cdot\boldsymbol{S}_j$ is that $E_M=-NzJ\expval{\boldsymbol{S}_i\cdot\boldsymbol{S}_j}$, it is reasonable to defines the magnetic specific heat as~\cite{Argyle1967}
\begin{equation}
    C_M=-NzJ\frac{\partial\expval{\boldsymbol{S}_i\cdot\boldsymbol{S}_j}}{\partial T}.
\end{equation}
As a result the deviation of local spin coherence due to the small change of local temperature $\theta$ is $C_M\theta=-NzJ\expval{\boldsymbol{S}_i\cdot\boldsymbol{S}_j}$ and the total volume change can be succinctly expressed into the form
\begin{equation}
    \epsilon_i^i=\alpha_E\theta+\beta_T\gamma_M\frac{C_M}{V}\theta=(\alpha_E+\alpha_M)\theta\equiv\Tilde{\alpha}\,\theta.
\end{equation}
We see that by merging the magnetoelastic coupling into the free energy, the thermal expansion coefficient $\alpha_E$ should be extended to the one including the magnetic contribution $\Tilde{\alpha}=\alpha_E+\alpha_M$. 

\begin{figure}
    \centering
    \includegraphics[width=0.8\columnwidth]{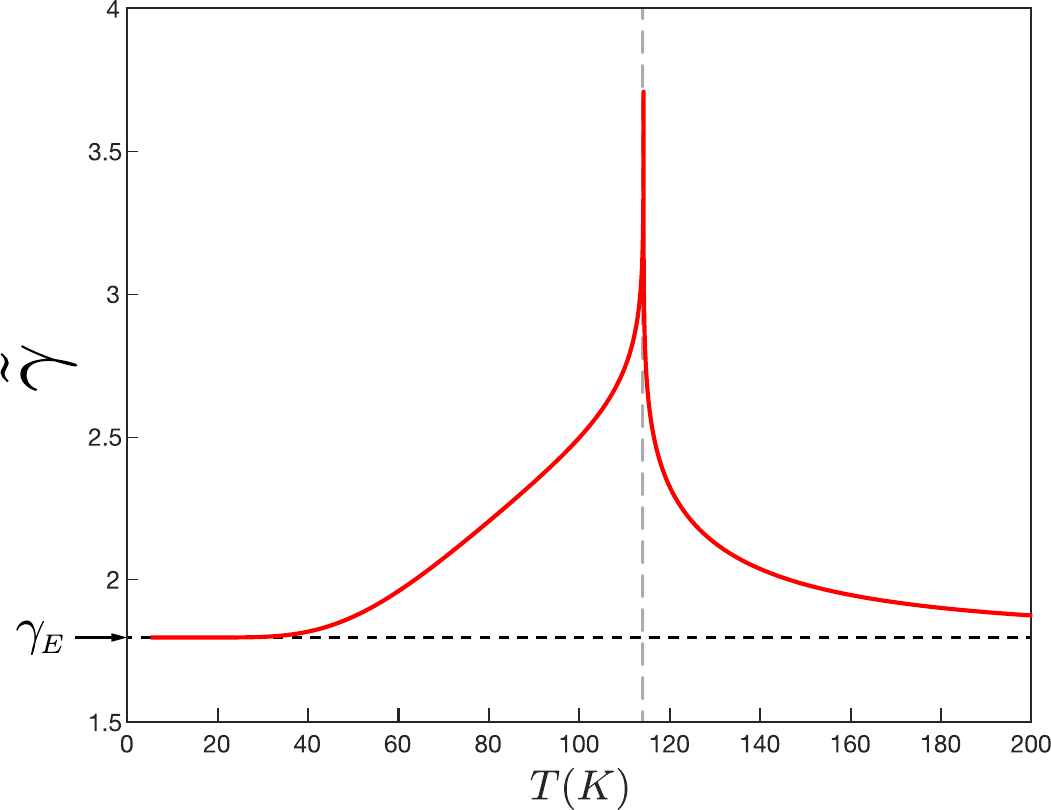}
    \caption[Temperature dependence for the effective Gr\"uneisen parameter $\Tilde{\gamma}$.]{Temperature dependence for the effective Gr\"uneisen parameter $\Tilde{\gamma}$ derived from Eq.~\ref{eqn:gamma_eff}. The elastic parameter is calculated to be to be $\gamma_E=1.798$~\cite{Makars2020} and the ratio is chosen to be $\gamma_M\big/\gamma_E=\nu=4$. The $\Tilde{\gamma}$ start from $\gamma_E$ because the $C_M\approx 0$ for small temperatures.}
    \label{fig:gamma}
\end{figure}

The magnetic Gr\"uneisen relation $\alpha_M=\beta_T\rho\gamma_MC_M$ is almost similar to the elastic counterpart ($\alpha_E=\beta_T\gamma_E\rho C_V$) meaning the thermal and magnetic properties both originate from the variation of spin coherence and it is the magnetic Gr\"uneisen parameter makes them a connection. Therefore the overall thermal expansion coefficient for the hybrid system can be written into the form
\begin{equation}
    \begin{aligned}
        \Tilde{\alpha}&=\beta_T\rho\gamma_EC_E+\beta_T\rho\gamma_MC_M=\beta_T\rho\left(\gamma_EC_E+\gamma_MC_M\right)\\
        &=\beta_T\rho\Tilde{\gamma}C_V,
    \end{aligned}
    \label{eqn:alpha_plate}
\end{equation}
which maintains the Gr\"uneisen relation formalism but with $C_V=C_E+C_M$ is the total specific heat combining the elastic and magnetic ones and with the effective Gr\"uneisen parameter defined as
\begin{equation}
    \Tilde{\gamma}=\frac{\gamma_EC_E+\gamma_MC_M}{C_E+C_M}.
    \label{eqn:gamma_eff}
\end{equation}
Although the elastic and magnetic Gr\"uneisen parameters are both almost independent of temperature~\cite{Shapira1976,McWhan1966}, the effective Gr\"uneisen parameter usually presents a peak at phase transition $T_N$ (as shown in Fig.~\ref{fig:gamma}). This phenomenon originates from the anomaly of magnetic specific heat near phase transition rendering the $\Tilde{\gamma}\approx\gamma_E$ for $T$ far away from $T_N$ and $\Tilde{\gamma}\approx\gamma_M$ for the $T$ close to $T_N$. Usually the $\gamma_M$ is several times larger than the elastic $\gamma_E$ and it can be theoretically predicted based on detailed study of magnetic structure~\cite{Gomes2019}. In this work, however, we shall simplify the analysis by assuming a phenomenological factor $\nu=\gamma_M\big/\gamma_E$ which can be further determined by fitting the theoretical prediction of the thermal observables such as the $\Tilde{\alpha}$ and $Q^{-1}$ to the measured values.

In this way the part of thermal expansion mediated by magnetostriction can be effectively absorbed into the non-magnetic formalism simply by replacing $\alpha_E$ with $\Tilde{\alpha}$. Together, the specific heat and thermal conductivity in the elastic and thermal dynamics equation (Eq.~\ref{eqn:2Dplate_elastic} and Eq.~\ref{eqn:2Dplate_thermal}) should also be replaced by the total specific heat $C_V=C_E+C_M$ and total thermal conductivity $\kappa=\kappa_E+\kappa_M$~\cite{Prasai2017}. The overall damping coefficient $Q^{-1}$ for the elastic and magnetic hybrid plate has the form (Eq.~\ref{eqn:Qinverse_plate_nomag})
\begin{widetext}
    \begin{equation}
        Q^{-1}=\frac{1+\sigma}{1-\sigma}\frac{Y\Tilde{\alpha}^2 T}{\rho C_V}\left(\frac{6}{\xi^{2}}-\frac{6}{\xi^{3}}\frac{\sinh\xi+\sin\xi}{\cosh\xi+\cos\xi}\right),\quad \xi=h\sqrt{\frac{\omega_0\rho C_V}{2\kappa}},
    \label{eqn:Qinverse_plate}
    \end{equation}    
\end{widetext}
with the $\Tilde{\alpha}$, $C_V$, and $\kappa$ are thermal observables which can be measured and predicted based on the theory developed in this chapter.

\section{Model validation through the thermal observables measured for the 2D AFM material FePS3}
\label{sec:model_validation}

\begin{figure}
    \centering
    \includegraphics[width=\columnwidth]{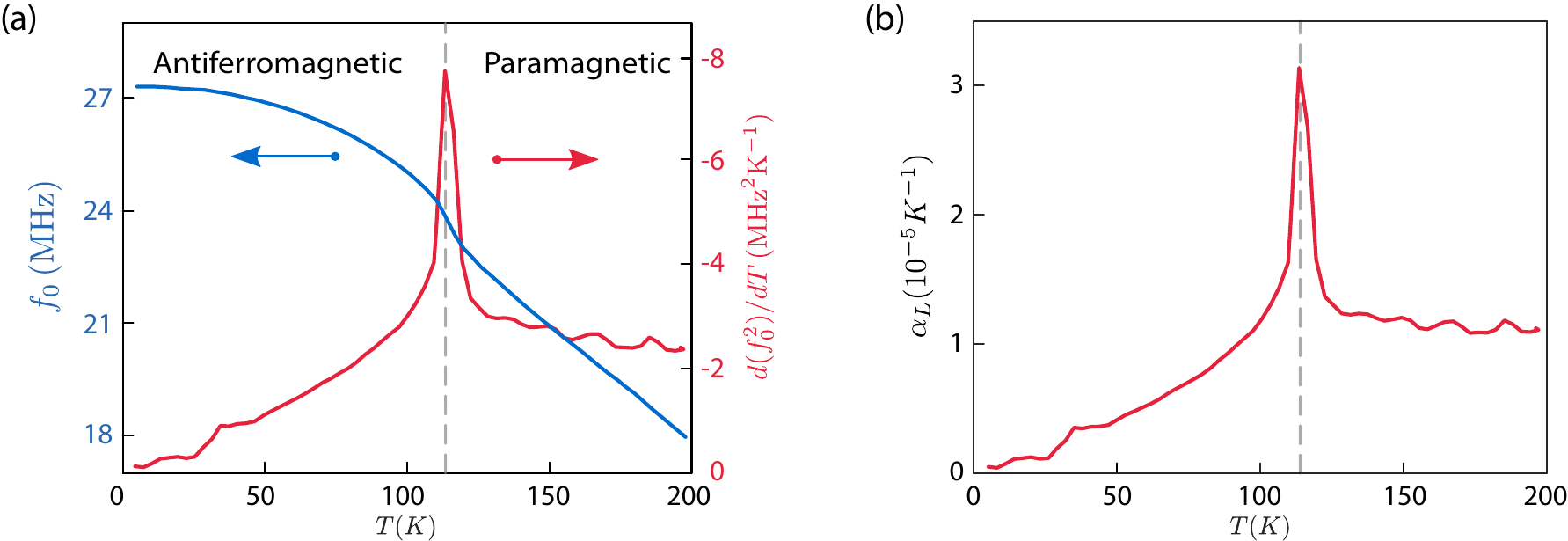}
    \caption[Measured decreasing of fundamental frequency for the FePS\textsubscript{3} resonator and the derived linear expansion coefficient.]{(a) Solid-blue line: the measured fundamental resonator frequency $f_0$ as function of temperature for the FePS\textsubscript{3} plate-membranes. Solid-red line: the derivative of $f_0^2$ to $T$. (b) Derived linear thermal expansion coefficient of FePS\textsubscript{3} plate-membranes according to Eq.~\ref{eqn:alpha_makars}. Quoted from the Fig.2 in Ref.~\cite{Makars2020}.}
    \label{fig:makars_alpha}
\end{figure}

Here we validate the theory developed in this chapter by calculating the linear thermal expansion coefficient $\alpha_L$ and damping factor $Q^{-1}$ of the Ising-type 2D antiferromagnetic material FePS\textsubscript{3} whose phase transition temperature is about $T_N=114K$~\cite{Makars2020}. In the published paper (Ref.~\cite{Makars2020}), \u{S}i\u{s}kins and etc. have measured the vibration frequency of the base model $f_0$ for the membrane-plate of FePS\textsubscript{3} in the setup of Fig.~\ref{fig:makars_setup}. According to Ref.~\cite{Andres2013} the resonance frequency of the round resonator in the membrane-plate regime can be approximated by
\begin{equation}
    f_0=\sqrt{f_{\mathrm{membrane}}^2+f_{\mathrm{plate}}^2}\,,
\end{equation}
in which the plate frequency is $f_{\mathrm{plate}}=\omega_0\big/2\pi$ according to the Eq.~\ref{eqn:frequency_plate} and the membranes fundamental frequency is 
\begin{equation}
    f_{\mathrm{membranes}}=\frac{2.4048}{2\pi a}\sqrt{\frac{N}{\rho h}},
\end{equation}
with $N=N_0+Yh\epsilon_r^\mathrm{th}\big/(1-\sigma)$ is the in-plane tension along radial direction. $N_0$ is the initial tension introduced by fabrication and can be further tuned by the external gate voltage $V_G$. The second part comes from the thermal expansion of the membranes which becomes a solely factor for the temperature dependency of the resonance frequency $f_0$ if we assume the plate frequency $f_\mathrm{plate}$ is independent to the environmental temperature because the Young's modulus $Y$ and Poisson coefficient $\sigma$ are almost independent to a small range of $T$ varying from $0$ to $200\,K$ in the \u{S}i\u{s}kins' experiment. The thermal strain is relate to the linear expansion coefficient of the resonator and silicon substrate by the relation $d\epsilon_r^\mathrm{th}\big/dT=-(\alpha_L-\alpha_\mathrm{si})$. As a consequence, by measuring the temperature dependency of $f_0(T)$ one can derive the thermal expansion coefficient of FePS\textsubscript{3} such that
\begin{equation}
    \alpha_L=\alpha_\mathrm{si}-\left(\frac{2\pi a}{2.4048}\right)^2\frac{\rho(1-\sigma)}{Y}\frac{d(f_0^2)}{dT}.
    \label{eqn:alpha_makars}
\end{equation}
The experimental measurement are presented in Fig.~\ref{fig:makars_alpha} and one indeed observes the $\alpha_L$ anomaly around the phase transition.

\begin{figure}
    \centering
    \includegraphics[width=0.8\columnwidth]{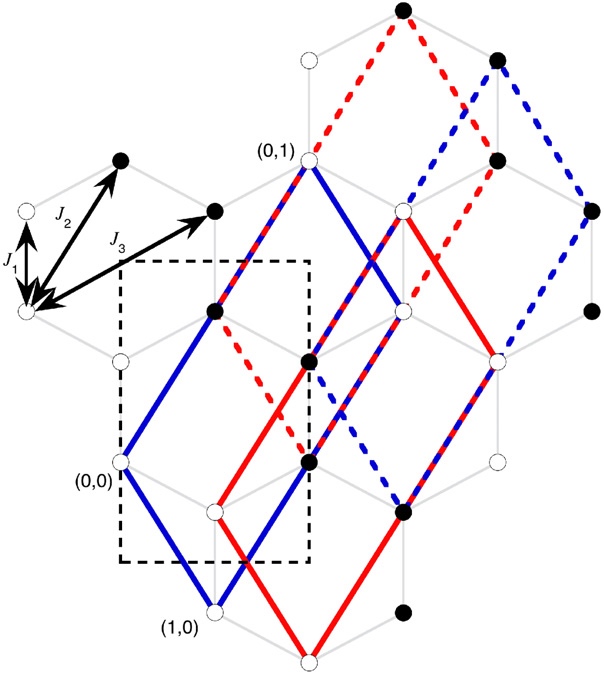}
    \caption[Schematic of the magnetic lattice for FePS\textsubscript{3} with the exchange coupling constants $J_1$, $J_2$, and $J_3$ indicated.]{Schematic of the magnetic lattice for FePS\textsubscript{3} quoted from Ref.~\cite{Wildes2012}. White dots mean the spin pointing out of the page and the black dots mean the spins pointing into the page. $J_1,J_2,J_3$ are the first-, second-, and third nearest neighbour interaction for the Hamiltonian $H=-\sum_{i,j} J_{i,j}\boldsymbol{S}_i\cdot\boldsymbol{S}_j$~\cite{Lancon2016}. The magnon dispersion relation with the effective exchange field is calculated based on the sub-lattice structure indicated by the red and blue rhombus. Total spin of magnetic Fe atom is $S=2$ and the coordination number for sublattice is $z=2$.}
    \label{fig:magnetic_FePS3}
\end{figure}

\begin{figure}
    \centering
    \includegraphics[width=\columnwidth]{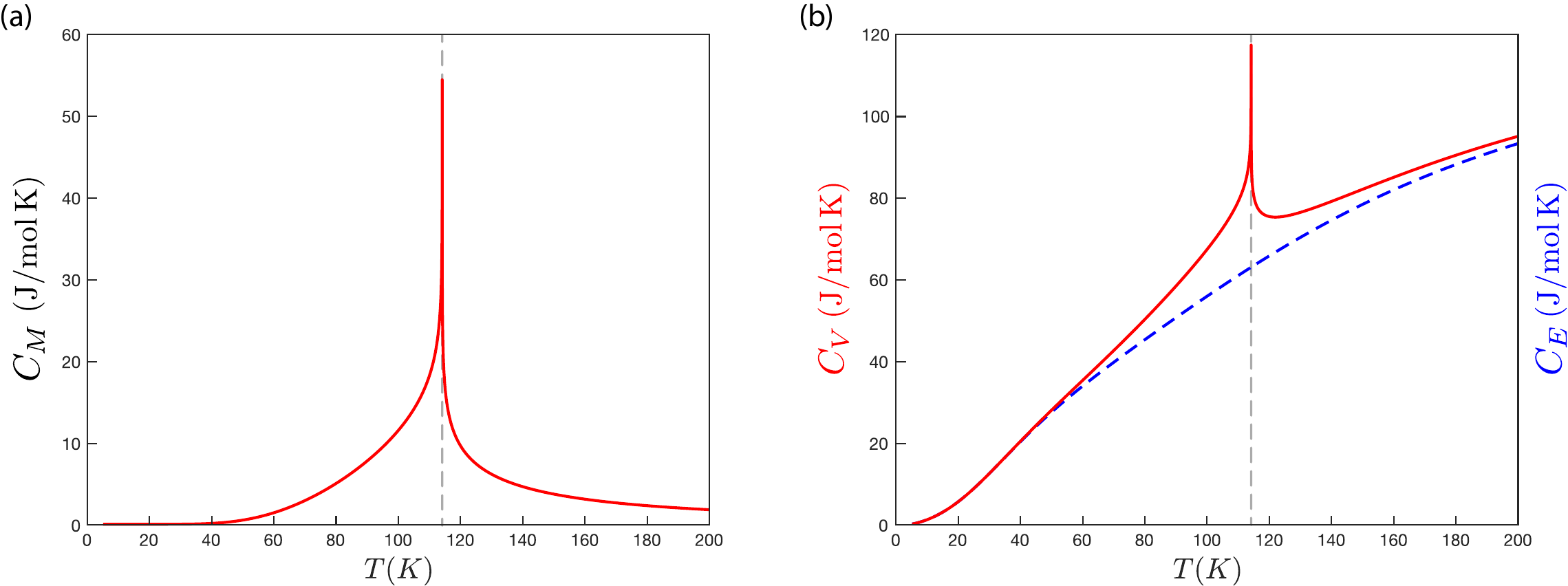}
    \caption[Theoretical predictions for the magnetic and total specific heat for the FePS\textsubscript{3} membranes.]{(a) Magnetic specific heat $C_M=C_\mathrm{Is}+C_\mathrm{mag}$ is the sum of the 2D Ising statistics and the magnon's contribution. (b) Solid red: total specific heat $C_V=C_E+C_M$ of the FePS\textsubscript{3}. It shows anomaly around the phase transition because the divergence of magnetic $C_M$. Doted blue: the elastic specific heat $C_E=(1-0.54)C_\mathrm{db}+0.54C_\mathrm{ei}$ according to Ref.~\cite{takano2004}. We point out that there are $5\,\mathrm{mol}$ of atoms per $1\,\mathrm{mol}$ molecule for the FePS\textsubscript{3} compound.}
    \label{fig:specific_heat}
\end{figure}

From the theoretical point of view, the linear expansion coefficient is one-third of the volume expansion coefficient developed in the previous section of the hybrid system, namely $\alpha_L=\Tilde{\alpha}/3$ based on Eq.~\ref{eqn:alpha_plate}. In order to derive the theoretical prediction of $\alpha_L$, one needs to calculate the specific heat of the elastic and magnetic parts. Firstly, for the magnetic specific of the Ising origin (Eq.~\ref{eqn:cm_honey}), the effective coupling energy $J'$ is derived from the measured Neel temperature $T_N=114K$ and according to Eq.~\ref{eqn:Neel} we have $J'=6.48\,\mathrm{meV}$. Therefore the nearest neighbour spin-to-spin coupling energy in the Hamiltonian $H=-2J\sum\boldsymbol{S}_i\cdot\boldsymbol{S}_j$ has the value $J=J'\big/2S^2=0.81\,\mathrm{meV}$ since the atomic spin for FePS\textsubscript{3} is $S=2$. One sees that the derived $J$ is very close to the first-nearest neighbour interaction (shown in Fig.~\ref{fig:magnetic_FePS3}) $J_1=2J\approx 1.5\,\mathrm{meV}$ measured in the neutron scattering experiment~\cite{Wildes2012,Lancon2016}. Using this derived $J'$ we plot the $C_\mathrm{Is}$ in Fig.~\ref{fig:C_Ising}(b). Secondly, for the magnetic specific of the magnon origin (Eq.~\ref{eqn:cm_2D}), it is necessary to figure out the exchange and anisotropy field on the sublattices in order to apply the dispersion relation in Eq.~\ref{eqn:2Dmagnon}. However, according to the magnetostriction effect the inter-atomic interaction are modulated by the strain and varies with the membranes thickness~\cite{Jianmin2021}. Here we simplify the analysis by selecting the effective field as $\mu_0 H_E=69\,\mathrm{Tesla}$ and $\mu_0 H_A=138\,\mathrm{Tesla}$ in order to best fit the derived $C_M(T)$ and $\alpha_L(T)$ with the measured data. According to the relation $H_E=2\abs{J}zS/\mu_0\gamma$, the effective interaction between sublattices then becomes $J_\mathrm{sub}\approx-1\,\mathrm{meV}$ and anisotropy is $A\approx6\,\mathrm{meV}$ which are close to the measured data whose values take $J_2=-0.04\,\mathrm{meV}$, $J_3=-0.96\,\mathrm{meV}$, and $A=3.78\,\mathrm{meV}$ as quoted from Refs.~\cite{Lancon2016,Wildes2012}. The calculated $C_\mathrm{mag}$ is shown in Fig.~\ref{fig:integral_compare}(a) and the total magnetic specific heat $C_M$ is shown in Fig.~\ref{fig:specific_heat}(a).
Obtained from first-principle calculation, the elastic parameters of FePS\textsubscript{3} are $Y=103\,\mathrm{GPa}$, $\sigma=0.304$, $\rho=3375\,\mathrm{kg}\,\mathrm{m}^{-3}$ and $\Bar{v}=3823\,\mathrm{m}\,\mathrm{s}^{-1}$~\cite{Makars2020}. According to the Ref.~\cite{takano2004}, the elastic specific heat for FePS\textsubscript{3} is a mixing of Debye and Einstein parts with the Debye temperature $T_\mathrm{db}=236\,\mathrm{K}$ and Einstein temperature $T_\mathrm{ei}=523\,\mathrm{K}$. The suggested combination ratio is $0.54$ and the elastic specific heat $C_E=(1-0.54)C_\mathrm{db}+0.54C_\mathrm{ei}$ can be derived from Eq.~\ref{eqn:cph_2D}. In Fig.~\ref{fig:specific_heat}(b) we present the calculated $C_E$ as doted blue line and the total specific heat $C_V=C_E+C_M$ as solid red line.
\begin{figure}
    \centering
    \includegraphics[width=\columnwidth]{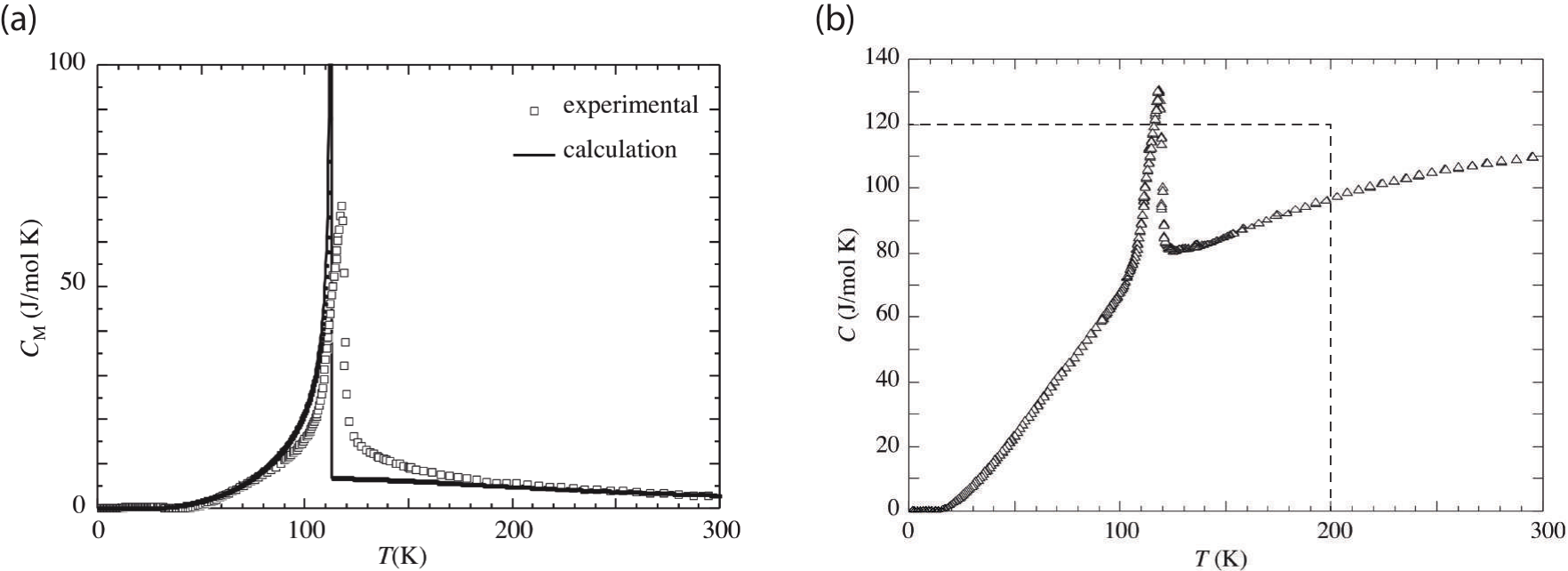}
    \caption[Experimentally measured magnetic and total specific heat for the FePS\textsubscript{3} compound.]{Measured Specific heat for FePS\textsubscript{3} quoted from Takano's paper~\cite{takano2004}. (a) the experimental data and Takano's prediction for $C_M$. In his calculation, the magnetic specific heat instantly decays to zero which does not fits into the measurements whereas my plotting fits better. (b) the experimental data for the total specific heat. Note here the temperature ranges from $0$ to $300\,\mathrm{K}$ while in my plotting the temperature stops at $200\,\mathrm{K}$.}
    \label{fig:takano}
\end{figure}

\begin{figure}[htb]
    \centering
    \includegraphics[width=0.8\columnwidth]{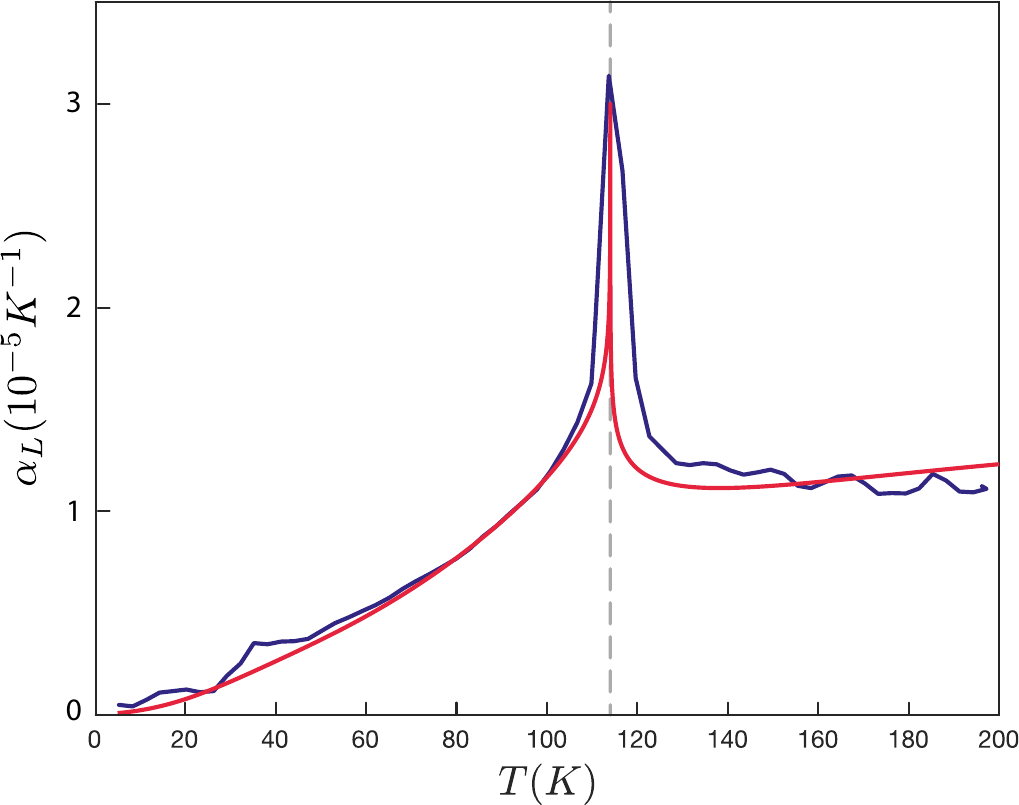}
    \caption[Comparison between theoretical prediction and measured data for the linear expansion coefficient of FePS\textsubscript{3} membranes.]{Solid red: theoretical predicted linear expansion coefficient $\alpha_L=\Tilde{\alpha}\big/3$ based on the Eq.~\ref{eqn:alpha_plate} with the derived specific heat from Fig.~\ref{fig:specific_heat} and effective Gr\"uneisen parameter $\Tilde{\gamma}$ from Fig.~\ref{fig:gamma}. Solid blue: experimental derived $\alpha_L$ from Eq.~\ref{eqn:alpha_makars}(b).}
    \label{fig:alpha}
\end{figure}

Our theoretical predictions fit well the measured data shown in Fig.~\ref{fig:takano} and therefore validate the choice of parameters and the applicability of our model. Furthermore, using these parameters we get the elastic Gr\"uneisen factor $\gamma_E=\frac{3}{2}\left(\frac{1+\sigma}{2-3\sigma}\right)=1.798$ and the compressibility $\beta_T=1.14\times10^{-11}\,\mathrm{Pa}^{-1}$. By assuming the ratio $\nu=\gamma_M\big/\gamma_E=4$ and applying the derived specific heats we calculate and plot the effective Gr\"uneisen parameter $\Tilde{\gamma}$ as function of temperature in Fig.~\ref{fig:gamma}. It is then straightforwards to derive the overall linear expansion coefficient for the hybrid system $\alpha_L=\Tilde{\alpha}\big/3$ based on equation~\ref{eqn:alpha_plate}. Bear in mind that if one uses molar specific heat from the Fig.~\ref{fig:specific_heat}(b), the density should also chosen to be the molar density which is $\rho=18443\,\mathrm{mol}\,\mathrm{m}^{-3}$ for FePS\textsubscript{3}. Showing in Fig.~\ref{fig:alpha} the theoretical prediction for $\alpha_L$ fits well the measured data which consolidates the scheme of merging the magnetoelastic coupling into the non-magnetic equation of motions for the hybrid system.

\begin{figure}[htb!]
    \centering
    \includegraphics[width=\columnwidth]{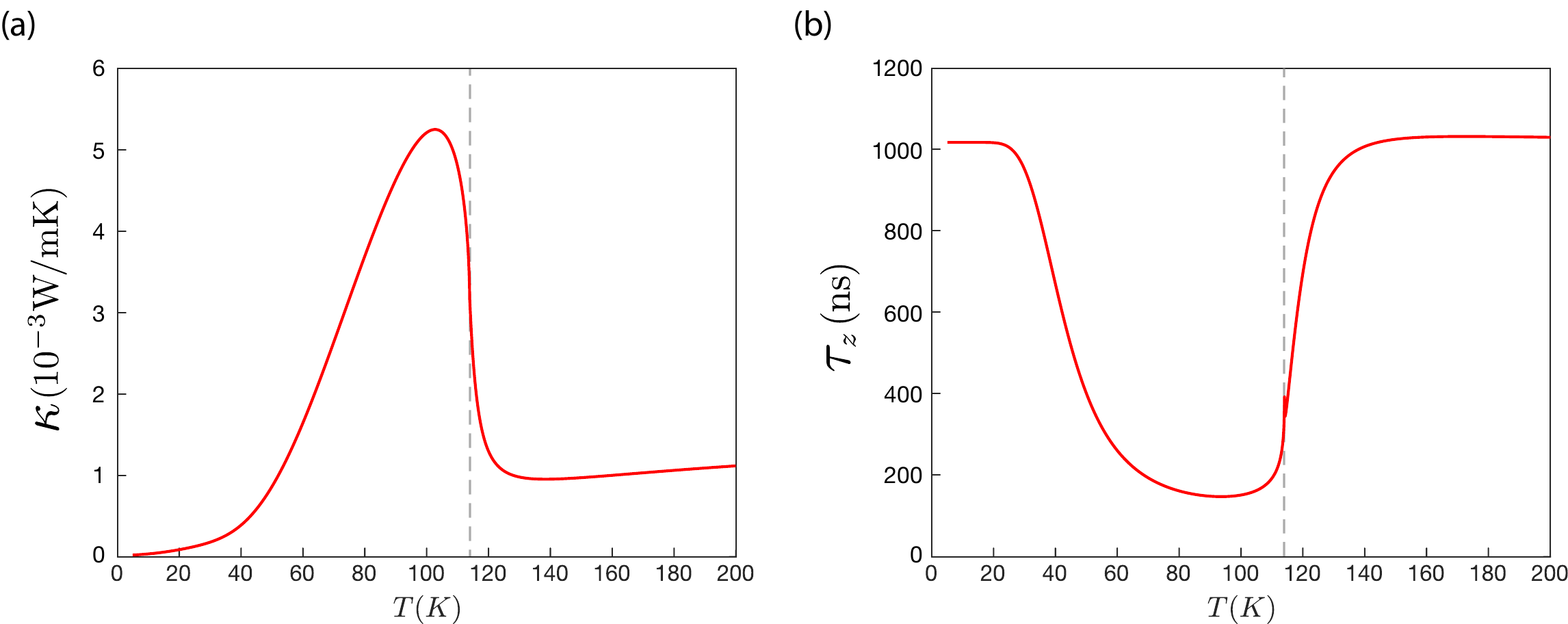}
    \caption[Theoretical derived thermal conductivity and thermal time constant for the FePS\textsubscript{3} membranes.]{(a) Total thermal conductivity which is a sum of the phonon and magnons' contribution as thermal carriers $\kappa=\kappa_\mathrm{ph}+\kappa_\mathrm{mag}$. For the membranes setup the magnetic part dominates before the phase transition and vanishes afterwards. (b) The thermal time constant along the $\hat{z}$-th direction $\tau_z=h^2\rho C_V\big/\pi\kappa$. The smaller of $\tau_z$ means the faster of temperature approaching equilibrium.}
    \label{fig:kappa_total}
\end{figure}

\begin{figure}[htb!]
    \centering
    \includegraphics[width=0.8\columnwidth]{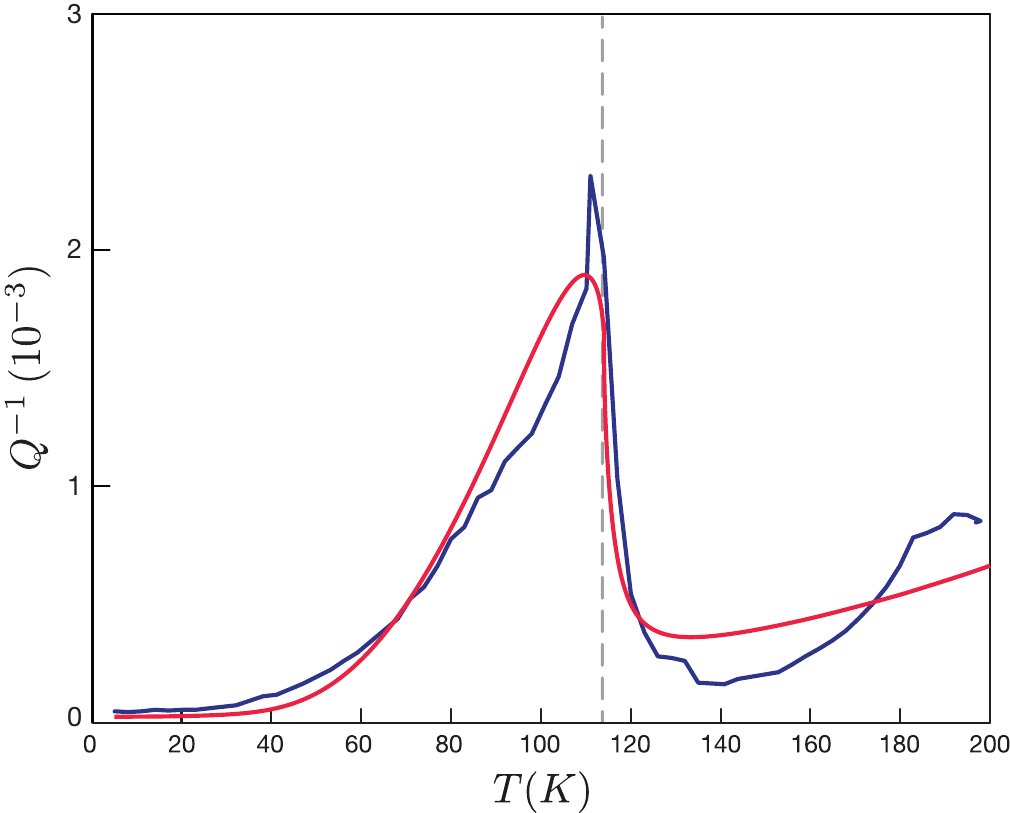}
    \caption[Comparison between theoretical prediction and measured data for the damping coefficient of FePS\textsubscript{3} resonator.]{Solid red: theoretical prediction for damping coefficient $Q^{-1}$ based on the Eq.~\ref{eqn:Qinverse_plate} with FePS\textsubscript{3} material parameters stated in the main texts. Solid blue: experimentally measured curve quoted from \u{S}i\u{s}kins paper (Ref.~\cite{Makars2020}).}
    \label{fig:Qinverse}
\end{figure}

In order to calculate and plot the damping coefficient $Q^{-1}$ according to Eq.~\ref{eqn:Qinverse_plate} one still needs to know the temperature dependence of thermal conductivity $\kappa$ especially in the hybrid materials whose thermal conduction has several different origins. As for the FePS\textsubscript{3}, we have $\kappa=\kappa_\mathrm{ph}+\kappa_\mathrm{mag}$ and we can ignore the scattering between phonons and magnons because the magnon's energy in antiferromagnetic is usually at the range of THz while the phonon's energy is usually of several GHz which means the coupling between these two quasi-particles is small. As stated in the previous section, particle lifetime is limited by the boundary scattering and can be treated as a constant $\tau=\tau_0$. The $\kappa_\mathrm{mag}$ can be derived according to Eq.~\ref{eqn:km_2D} together with material constants and the fitting parameter $\tau_{0,\mathrm{mag}}\approx3.8\,\mathrm{ps}$~\cite{Xufei2018}. As for the phonon's contribution, we simplify the analysis by utilizing the Debye averaged sound velocity and apply the fitting parameter $\tau_{0,\mathrm{ph}}\approx0.8\,\mathrm{ps}$ such that $\kappa_\mathrm{ph}=C_E\Bar{v}^2\tau_{0,\mathrm{ph}}$. The total thermal conductivity is plotted in Fig.~\ref{fig:kappa_total}(a) and we see it is much smaller than the measured value for bulk FePS\textsubscript{3} compound which has $\kappa\approx1\,\mathrm{W}\big/\mathrm{mK}$ at room temperature~\cite{Kargar2020}. This is due to the membranes geometry whose thickness is only $h=45\,\mathrm{nm}$ which limits mobility of phonons and thus the small thermal conductivity. The transverse thermal time constant $\tau_z=h^2\rho C_V\big/\pi\kappa$, which measures the time for establishing the temperature equilibrium across the plate, is also plotted and it is close to the \u{S}i\u{s}kins measurement. With the parameter $\xi=\pi\sqrt{f_0\,\tau_z}$ and based on the previously derived expansion coefficient $\Tilde{\alpha}$ and total specific heat $C_V$, we have the damping coefficient $Q^{-1}$ derived and shown in Fig.~\ref{fig:Qinverse}. We see the agreement between theoretical prediction and experiment data is good enough and the drop of thermal transfer after phase transition can be ascribed to the depletion of magnons as thermal carriers.

\section{Summary and outlook}
\label{sec:summary}

In conclusion we studied the magnetoelastic effect on the thermal transfer within the thin AFM plate for a wide range of temperature across the magnetic phase transition. In specific, we developed a theory of merging the exchange magnetoelastic interaction into the thermal elastic free energy and further predicted the temperature dependence for observables such as specific heat $C_V$, linear expansion coefficient $\Tilde{\alpha}$, and damping factor $Q^{-1}$ for the quasi-2D Ising AFM material FePS\textsubscript{3}. Compared to the experimentally measured data, our theoretical predictions agree very well especially for the specific heat and linear expansion coefficient. As for the transport related property, the theoretical plot of $Q^{-1}(T)$ shows the overall trend consistent with the measured curve and it still has rooms for improvement. It is because in this work we have simplified the magnon and phonon velocity $\boldsymbol{v}_k$ to be homogeneous and utilized an isotropic thermal conductivity for analysis. According to the quasi 2D material these assumptions may not sufficient enough and one can improve these transport properties by studying the detailed lattice structure~\cite{Wildes2012}. It may also helpful to find a double peak effect~\cite{Yoichi1998} for the $\kappa(T)$ is helpful to explain the secondary surging of $Q^{-1}$ after $T>T_N$. However, our theoretical treatment builds a general scheme to study the thermal observables for the magnetic-elastic-thermal integrated system. The key is generalizing the Gr\"uneisen relation by incorporating various contributions and arriving at an effective Gr\"uneisen coefficient $\Tilde{\gamma}$ (Eq.~\ref{eqn:gamma_eff}). This quantity essentially describes the variation of internal energy with respect to the volume change and its temperature dependency represents the changing of \textit{weight} in the internal energy for each components in the hybrid system. Therefore the scheme developed in this chapter can be extended to include other contributors such as electrons in the spintronic and spin-caloritronic devices.

\bibliography{references}

\begin{thebibliography}{40}%
\makeatletter
\providecommand \@ifxundefined [1]{%
 \@ifx{#1\undefined}
}%
\providecommand \@ifnum [1]{%
 \ifnum #1\expandafter \@firstoftwo
 \else \expandafter \@secondoftwo
 \fi
}%
\providecommand \@ifx [1]{%
 \ifx #1\expandafter \@firstoftwo
 \else \expandafter \@secondoftwo
 \fi
}%
\providecommand \natexlab [1]{#1}%
\providecommand \enquote  [1]{``#1''}%
\providecommand \bibnamefont  [1]{#1}%
\providecommand \bibfnamefont [1]{#1}%
\providecommand \citenamefont [1]{#1}%
\providecommand \href@noop [0]{\@secondoftwo}%
\providecommand \href [0]{\begingroup \@sanitize@url \@href}%
\providecommand \@href[1]{\@@startlink{#1}\@@href}%
\providecommand \@@href[1]{\endgroup#1\@@endlink}%
\providecommand \@sanitize@url [0]{\catcode `\\12\catcode `\$12\catcode `\&12\catcode `\#12\catcode `\^12\catcode `\_12\catcode `\%12\relax}%
\providecommand \@@startlink[1]{}%
\providecommand \@@endlink[0]{}%
\providecommand \url  [0]{\begingroup\@sanitize@url \@url }%
\providecommand \@url [1]{\endgroup\@href {#1}{\urlprefix }}%
\providecommand \urlprefix  [0]{URL }%
\providecommand \Eprint [0]{\href }%
\providecommand \doibase [0]{https://doi.org/}%
\providecommand \selectlanguage [0]{\@gobble}%
\providecommand \bibinfo  [0]{\@secondoftwo}%
\providecommand \bibfield  [0]{\@secondoftwo}%
\providecommand \translation [1]{[#1]}%
\providecommand \BibitemOpen [0]{}%
\providecommand \bibitemStop [0]{}%
\providecommand \bibitemNoStop [0]{.\EOS\space}%
\providecommand \EOS [0]{\spacefactor3000\relax}%
\providecommand \BibitemShut  [1]{\csname bibitem#1\endcsname}%
\let\auto@bib@innerbib\@empty
\bibitem [{\citenamefont {Gong}\ and\ \citenamefont {Zhang}(2019)}]{Cheng2019}%
  \BibitemOpen
  \bibfield  {author} {\bibinfo {author} {\bibfnamefont {C.}~\bibnamefont {Gong}}\ and\ \bibinfo {author} {\bibfnamefont {X.}~\bibnamefont {Zhang}},\ }\bibfield  {title} {\bibinfo {title} {Two-dimensional magnetic crystals and emergent heterostructure devices},\ }\bibfield  {journal} {\bibinfo  {journal} {Science}\ }\textbf {\bibinfo {volume} {363}},\ \href {https://doi.org/10.1126/science.aav4450} {10.1126/science.aav4450} (\bibinfo {year} {2019})\BibitemShut {NoStop}%
\bibitem [{\citenamefont {Gong}\ \emph {et~al.}(2017)\citenamefont {Gong}, \citenamefont {Li}, \citenamefont {Li}, \citenamefont {Ji}, \citenamefont {Stern}, \citenamefont {Xia},\ and\ \citenamefont {etc.}}]{Gong2017}%
  \BibitemOpen
  \bibfield  {author} {\bibinfo {author} {\bibfnamefont {C.}~\bibnamefont {Gong}}, \bibinfo {author} {\bibfnamefont {L.}~\bibnamefont {Li}}, \bibinfo {author} {\bibfnamefont {Z.}~\bibnamefont {Li}}, \bibinfo {author} {\bibfnamefont {H.}~\bibnamefont {Ji}}, \bibinfo {author} {\bibfnamefont {A.}~\bibnamefont {Stern}}, \bibinfo {author} {\bibfnamefont {Y.}~\bibnamefont {Xia}},\ and\ \bibinfo {author} {\bibnamefont {etc.}},\ }\bibfield  {title} {\bibinfo {title} {Discovery of intrinsic ferromagnetism in two-dimensional van der waals crystals},\ }\href {https://doi.org/https://doi.org/10.1038/nature22060} {\bibfield  {journal} {\bibinfo  {journal} {Nature}\ }\textbf {\bibinfo {volume} {546}},\ \bibinfo {pages} {265} (\bibinfo {year} {2017})}\BibitemShut {NoStop}%
\bibitem [{\citenamefont {Li}\ \emph {et~al.}(2019)\citenamefont {Li}, \citenamefont {Ruan},\ and\ \citenamefont {Zeng}}]{Hui2019}%
  \BibitemOpen
  \bibfield  {author} {\bibinfo {author} {\bibfnamefont {H.}~\bibnamefont {Li}}, \bibinfo {author} {\bibfnamefont {S.}~\bibnamefont {Ruan}},\ and\ \bibinfo {author} {\bibfnamefont {Y.-J.}\ \bibnamefont {Zeng}},\ }\bibfield  {title} {\bibinfo {title} {Intrinsic van der waals magnetic materials from bulk to the 2{D} limit: New frontiers of spintronics},\ }\href {https://doi.org/https://doi.org/10.1002/adma.201900065} {\bibfield  {journal} {\bibinfo  {journal} {Advanced Materials}\ }\textbf {\bibinfo {volume} {31}},\ \bibinfo {pages} {1900065} (\bibinfo {year} {2019})}\BibitemShut {NoStop}%
\bibitem [{\citenamefont {Boona}\ \emph {et~al.}(2014)\citenamefont {Boona}, \citenamefont {Myers},\ and\ \citenamefont {Heremans}}]{Boona2014}%
  \BibitemOpen
  \bibfield  {author} {\bibinfo {author} {\bibfnamefont {S.~R.}\ \bibnamefont {Boona}}, \bibinfo {author} {\bibfnamefont {R.~C.}\ \bibnamefont {Myers}},\ and\ \bibinfo {author} {\bibfnamefont {J.~P.}\ \bibnamefont {Heremans}},\ }\bibfield  {title} {\bibinfo {title} {Spin caloritronics},\ }\href {https://doi.org/10.1039/C3EE43299H} {\bibfield  {journal} {\bibinfo  {journal} {Energy Environ. Sci.}\ }\textbf {\bibinfo {volume} {7}},\ \bibinfo {pages} {885} (\bibinfo {year} {2014})}\BibitemShut {NoStop}%
\bibitem [{\citenamefont {Kargar}\ \emph {et~al.}(2020)\citenamefont {Kargar}, \citenamefont {Coleman}, \citenamefont {Ghosh}, \citenamefont {Lee}, \citenamefont {Balandin},\ and\ \citenamefont {etc.}}]{Kargar2020}%
  \BibitemOpen
  \bibfield  {author} {\bibinfo {author} {\bibfnamefont {F.}~\bibnamefont {Kargar}}, \bibinfo {author} {\bibfnamefont {E.~A.}\ \bibnamefont {Coleman}}, \bibinfo {author} {\bibfnamefont {S.}~\bibnamefont {Ghosh}}, \bibinfo {author} {\bibfnamefont {J.}~\bibnamefont {Lee}}, \bibinfo {author} {\bibfnamefont {A.}~\bibnamefont {Balandin}},\ and\ \bibinfo {author} {\bibnamefont {etc.}},\ }\bibfield  {title} {\bibinfo {title} {Phonon and thermal properties of quasi-two-dimensional {FePS}\textsubscript{3} and {MnPS}\textsubscript{3} antiferromagnetic semiconductors},\ }\href {https://doi.org/doi:10.1021/acsnano.9b09839} {\bibfield  {journal} {\bibinfo  {journal} {ACS Nano}\ }\textbf {\bibinfo {volume} {14}},\ \bibinfo {pages} {2424 } (\bibinfo {year} {2020})}\BibitemShut {NoStop}%
\bibitem [{\citenamefont {Lee}\ \emph {et~al.}(2016)\citenamefont {Lee}, \citenamefont {Lee}, \citenamefont {Ryoo}, \citenamefont {Kang}, \citenamefont {Kim}, \citenamefont {Kim}, \citenamefont {Park}, \citenamefont {Park},\ and\ \citenamefont {Cheong}}]{Lee2016}%
  \BibitemOpen
  \bibfield  {author} {\bibinfo {author} {\bibfnamefont {J.-U.}\ \bibnamefont {Lee}}, \bibinfo {author} {\bibfnamefont {S.}~\bibnamefont {Lee}}, \bibinfo {author} {\bibfnamefont {J.~H.}\ \bibnamefont {Ryoo}}, \bibinfo {author} {\bibfnamefont {S.}~\bibnamefont {Kang}}, \bibinfo {author} {\bibfnamefont {T.~Y.}\ \bibnamefont {Kim}}, \bibinfo {author} {\bibfnamefont {P.}~\bibnamefont {Kim}}, \bibinfo {author} {\bibfnamefont {C.-H.}\ \bibnamefont {Park}}, \bibinfo {author} {\bibfnamefont {J.-G.}\ \bibnamefont {Park}},\ and\ \bibinfo {author} {\bibfnamefont {H.}~\bibnamefont {Cheong}},\ }\bibfield  {title} {\bibinfo {title} {Ising-type magnetic ordering in atomically thin {FePS}\textsubscript{3}},\ }\href {https://doi.org/10.1021/acs.nanolett.6b03052} {\bibfield  {journal} {\bibinfo  {journal} {Nano Letters}\ }\textbf {\bibinfo {volume} {16}},\ \bibinfo {pages} {7433} (\bibinfo {year} {2016})}\BibitemShut {NoStop}%
\bibitem [{\citenamefont {Lan\ifmmode~\mbox{\c{c}}\else \c{c}\fi{}on}\ \emph {et~al.}(2016)\citenamefont {Lan\ifmmode~\mbox{\c{c}}\else \c{c}\fi{}on}, \citenamefont {Walker}, \citenamefont {Ressouche}, \citenamefont {Ouladdiaf}, \citenamefont {Rule}, \citenamefont {McIntyre}, \citenamefont {Hicks}, \citenamefont {R\o{}nnow},\ and\ \citenamefont {Wildes}}]{Lancon2016}%
  \BibitemOpen
  \bibfield  {author} {\bibinfo {author} {\bibfnamefont {D.}~\bibnamefont {Lan\ifmmode~\mbox{\c{c}}\else \c{c}\fi{}on}}, \bibinfo {author} {\bibfnamefont {H.~C.}\ \bibnamefont {Walker}}, \bibinfo {author} {\bibfnamefont {E.}~\bibnamefont {Ressouche}}, \bibinfo {author} {\bibfnamefont {B.}~\bibnamefont {Ouladdiaf}}, \bibinfo {author} {\bibfnamefont {K.~C.}\ \bibnamefont {Rule}}, \bibinfo {author} {\bibfnamefont {G.~J.}\ \bibnamefont {McIntyre}}, \bibinfo {author} {\bibfnamefont {T.~J.}\ \bibnamefont {Hicks}}, \bibinfo {author} {\bibfnamefont {H.~M.}\ \bibnamefont {R\o{}nnow}},\ and\ \bibinfo {author} {\bibfnamefont {A.~R.}\ \bibnamefont {Wildes}},\ }\bibfield  {title} {\bibinfo {title} {Magnetic structure and magnon dynamics of the quasi-two-dimensional antiferromagnet {FePS}\textsubscript{3}},\ }\href {https://doi.org/10.1103/PhysRevB.94.214407} {\bibfield  {journal} {\bibinfo  {journal} {Phys. Rev. B}\ }\textbf {\bibinfo {volume} {94}},\ \bibinfo {pages} {214407} (\bibinfo {year} {2016})}\BibitemShut
  {NoStop}%
\bibitem [{\citenamefont {Wildes}\ \emph {et~al.}(2012)\citenamefont {Wildes}, \citenamefont {Rule}, \citenamefont {Bewley}, \citenamefont {Enderle},\ and\ \citenamefont {Hicks}}]{Wildes2012}%
  \BibitemOpen
  \bibfield  {author} {\bibinfo {author} {\bibfnamefont {A.~R.}\ \bibnamefont {Wildes}}, \bibinfo {author} {\bibfnamefont {K.~C.}\ \bibnamefont {Rule}}, \bibinfo {author} {\bibfnamefont {R.~I.}\ \bibnamefont {Bewley}}, \bibinfo {author} {\bibfnamefont {M.}~\bibnamefont {Enderle}},\ and\ \bibinfo {author} {\bibfnamefont {T.~J.}\ \bibnamefont {Hicks}},\ }\bibfield  {title} {\bibinfo {title} {The magnon dynamics and spin exchange parameters of {FePS}\textsubscript{3}},\ }\href {https://doi.org/10.1088/0953-8984/24/41/416004} {\bibfield  {journal} {\bibinfo  {journal} {Journal of Physics: Condensed Matter}\ }\textbf {\bibinfo {volume} {24}},\ \bibinfo {pages} {416004} (\bibinfo {year} {2012})}\BibitemShut {NoStop}%
\bibitem [{\citenamefont {Šiškins}\ \emph {et~al.}(2020)\citenamefont {Šiškins}, \citenamefont {Lee}, \citenamefont {Mañas-Valero},\ and\ \citenamefont {et~al}}]{Makars2020}%
  \BibitemOpen
  \bibfield  {author} {\bibinfo {author} {\bibfnamefont {M.}~\bibnamefont {Šiškins}}, \bibinfo {author} {\bibfnamefont {M.}~\bibnamefont {Lee}}, \bibinfo {author} {\bibfnamefont {S.}~\bibnamefont {Mañas-Valero}},\ and\ \bibinfo {author} {\bibnamefont {et~al}},\ }\bibfield  {title} {\bibinfo {title} {Magnetic and electronic phase transitions probed by nanomechanical resonators},\ }\href {https://doi.org/https://doi.org/10.1038/s41467-020-16430-2} {\bibfield  {journal} {\bibinfo  {journal} {Nature Communication}\ }\textbf {\bibinfo {volume} {11}},\ \bibinfo {pages} {2698} (\bibinfo {year} {2020})}\BibitemShut {NoStop}%
\bibitem [{\citenamefont {Takano}\ \emph {et~al.}(2004)\citenamefont {Takano}, \citenamefont {Arai}, \citenamefont {Arai}, \citenamefont {Takahashi}, \citenamefont {Takase},\ and\ \citenamefont {Sekizawa}}]{takano2004}%
  \BibitemOpen
  \bibfield  {author} {\bibinfo {author} {\bibfnamefont {Y.}~\bibnamefont {Takano}}, \bibinfo {author} {\bibfnamefont {N.}~\bibnamefont {Arai}}, \bibinfo {author} {\bibfnamefont {A.}~\bibnamefont {Arai}}, \bibinfo {author} {\bibfnamefont {Y.}~\bibnamefont {Takahashi}}, \bibinfo {author} {\bibfnamefont {K.}~\bibnamefont {Takase}},\ and\ \bibinfo {author} {\bibfnamefont {K.}~\bibnamefont {Sekizawa}},\ }\bibfield  {title} {\bibinfo {title} {Magnetic properties and specific heat of {MPS}\textsubscript{3} ({M}={M}n, {F}e, {Z}n)},\ }\href {https://doi.org/https://doi.org/10.1016/j.jmmm.2003.12.621} {\bibfield  {journal} {\bibinfo  {journal} {Journal of Magnetism and Magnetic Materials}\ }\textbf {\bibinfo {volume} {272-276}},\ \bibinfo {pages} {E593 } (\bibinfo {year} {2004})}\BibitemShut {NoStop}%
\bibitem [{\citenamefont {Landau}\ and\ \citenamefont {Lifshitz}(1986)}]{Landau1986}%
  \BibitemOpen
  \bibfield  {author} {\bibinfo {author} {\bibfnamefont {L.~D.}\ \bibnamefont {Landau}}\ and\ \bibinfo {author} {\bibfnamefont {E.~M.}\ \bibnamefont {Lifshitz}},\ }\bibfield  {title} {\bibinfo {title} {Chapter ii - the equilibrium of rods and plates},\ }in\ \href {https://doi.org/https://doi.org/10.1016/B978-0-08-057069-3.50009-7} {\emph {\bibinfo {booktitle} {Theory of Elasticity}}}\ (\bibinfo  {publisher} {Butterworth-Heinemann},\ \bibinfo {year} {1986})\ \bibinfo {edition} {third edition}\ ed.,\ pp.\ \bibinfo {pages} {38--86}\BibitemShut {NoStop}%
\bibitem [{\citenamefont {Huang}(2005)}]{Huang2005}%
  \BibitemOpen
  \bibfield  {author} {\bibinfo {author} {\bibfnamefont {C.-H.}\ \bibnamefont {Huang}},\ }\bibfield  {title} {\bibinfo {title} {Transverse vibration analysis and measurement for the piezoceramic annular plate with different boundary conditions},\ }\href {https://doi.org/https://doi.org/10.1016/j.jsv.2004.05.034} {\bibfield  {journal} {\bibinfo  {journal} {Journal of Sound and Vibration}\ }\textbf {\bibinfo {volume} {283}},\ \bibinfo {pages} {665} (\bibinfo {year} {2005})}\BibitemShut {NoStop}%
\bibitem [{\citenamefont {Sun}\ and\ \citenamefont {Tohmyoh}(2008)}]{Sun2008}%
  \BibitemOpen
  \bibfield  {author} {\bibinfo {author} {\bibfnamefont {Y.}~\bibnamefont {Sun}}\ and\ \bibinfo {author} {\bibfnamefont {H.}~\bibnamefont {Tohmyoh}},\ }\bibfield  {title} {\bibinfo {title} {Thermoelastic damping of the axisymmetric vibration of circular plate resonators},\ }\href {https://doi.org/https://doi.org/10.1016/j.jsv.2008.06.017} {\bibfield  {journal} {\bibinfo  {journal} {Journal of Sound and Vibration}\ }\textbf {\bibinfo {volume} {319}},\ \bibinfo {pages} {392} (\bibinfo {year} {2008})}\BibitemShut {NoStop}%
\bibitem [{\citenamefont {McPherson}(2019)}]{McPherson2019}%
  \BibitemOpen
  \bibfield  {author} {\bibinfo {author} {\bibfnamefont {J.~W.}\ \bibnamefont {McPherson}},\ }\href {https://doi.org/https://doi.org/10.1007/978-3-319-93683-3_18} {\emph {\bibinfo {title} {Heat Generation and Dissipation}}}\ (\bibinfo  {publisher} {Springer International Publishing},\ \bibinfo {year} {2019})\BibitemShut {NoStop}%
\bibitem [{\citenamefont {Lifshitz}\ and\ \citenamefont {Roukes}(2000)}]{Lifshitz2000}%
  \BibitemOpen
  \bibfield  {author} {\bibinfo {author} {\bibfnamefont {R.}~\bibnamefont {Lifshitz}}\ and\ \bibinfo {author} {\bibfnamefont {M.~L.}\ \bibnamefont {Roukes}},\ }\bibfield  {title} {\bibinfo {title} {Thermoelastic damping in micro- and nanomechanical systems},\ }\href {https://doi.org/10.1103/PhysRevB.61.5600} {\bibfield  {journal} {\bibinfo  {journal} {Phys. Rev. B}\ }\textbf {\bibinfo {volume} {61}},\ \bibinfo {pages} {5600} (\bibinfo {year} {2000})}\BibitemShut {NoStop}%
\bibitem [{\citenamefont {Rezende}\ \emph {et~al.}(2014{\natexlab{a}})\citenamefont {Rezende}, \citenamefont {Rodr\'{\i}guez-Su\'arez}, \citenamefont {Lopez~Ortiz},\ and\ \citenamefont {Azevedo}}]{Rezende2014}%
  \BibitemOpen
  \bibfield  {author} {\bibinfo {author} {\bibfnamefont {S.~M.}\ \bibnamefont {Rezende}}, \bibinfo {author} {\bibfnamefont {R.~L.}\ \bibnamefont {Rodr\'{\i}guez-Su\'arez}}, \bibinfo {author} {\bibfnamefont {J.~C.}\ \bibnamefont {Lopez~Ortiz}},\ and\ \bibinfo {author} {\bibfnamefont {A.}~\bibnamefont {Azevedo}},\ }\bibfield  {title} {\bibinfo {title} {Thermal properties of magnons and the spin seebeck effect in yttrium iron garnet/normal metal hybrid structures},\ }\href {https://doi.org/10.1103/PhysRevB.89.134406} {\bibfield  {journal} {\bibinfo  {journal} {Phys. Rev. B}\ }\textbf {\bibinfo {volume} {89}},\ \bibinfo {pages} {134406} (\bibinfo {year} {2014}{\natexlab{a}})}\BibitemShut {NoStop}%
\bibitem [{\citenamefont {Shen}(2018)}]{Shen2018}%
  \BibitemOpen
  \bibfield  {author} {\bibinfo {author} {\bibfnamefont {K.}~\bibnamefont {Shen}},\ }\bibfield  {title} {\bibinfo {title} {Finite temperature magnon spectra in yttrium iron garnet from a mean field approach in a tight-binding model},\ }\href {https://doi.org/10.1088/1367-2630/aab951} {\bibfield  {journal} {\bibinfo  {journal} {New Journal of Physics}\ }\textbf {\bibinfo {volume} {20}},\ \bibinfo {pages} {043025} (\bibinfo {year} {2018})}\BibitemShut {NoStop}%
\bibitem [{\citenamefont {Simon}(2016)}]{Simon2016}%
  \BibitemOpen
  \bibfield  {author} {\bibinfo {author} {\bibfnamefont {S.~H.}\ \bibnamefont {Simon}},\ }\href {https://global.oup.com/academic/product/the-oxford-solid-state-basics-9780199680771?cc=us&lang=en&} {\emph {\bibinfo {title} {The Oxford Solid State Basics}}}\ (\bibinfo  {publisher} {Oxford University Press},\ \bibinfo {year} {2016})\BibitemShut {NoStop}%
\bibitem [{\citenamefont {L\'opez~Ortiz}\ \emph {et~al.}(2014)\citenamefont {L\'opez~Ortiz}, \citenamefont {Fonseca~Guerra}, \citenamefont {Machado},\ and\ \citenamefont {Rezende}}]{Ortiz2014}%
  \BibitemOpen
  \bibfield  {author} {\bibinfo {author} {\bibfnamefont {J.~C.}\ \bibnamefont {L\'opez~Ortiz}}, \bibinfo {author} {\bibfnamefont {G.~A.}\ \bibnamefont {Fonseca~Guerra}}, \bibinfo {author} {\bibfnamefont {F.~L.~A.}\ \bibnamefont {Machado}},\ and\ \bibinfo {author} {\bibfnamefont {S.~M.}\ \bibnamefont {Rezende}},\ }\bibfield  {title} {\bibinfo {title} {Magnetic anisotropy of antiferromagnetic {RbMnF}\textsubscript{3}},\ }\href {https://doi.org/10.1103/PhysRevB.90.054402} {\bibfield  {journal} {\bibinfo  {journal} {Phys. Rev. B}\ }\textbf {\bibinfo {volume} {90}},\ \bibinfo {pages} {054402} (\bibinfo {year} {2014})}\BibitemShut {NoStop}%
\bibitem [{\citenamefont {Cole}\ and\ \citenamefont {Ince}(1966)}]{Cole1966}%
  \BibitemOpen
  \bibfield  {author} {\bibinfo {author} {\bibfnamefont {P.~H.}\ \bibnamefont {Cole}}\ and\ \bibinfo {author} {\bibfnamefont {W.~J.}\ \bibnamefont {Ince}},\ }\bibfield  {title} {\bibinfo {title} {Equilibrium spin configuration and resonance behavior of {R}b{M}n{F}3},\ }\href {https://doi.org/10.1103/PhysRev.150.377} {\bibfield  {journal} {\bibinfo  {journal} {Phys. Rev.}\ }\textbf {\bibinfo {volume} {150}},\ \bibinfo {pages} {377} (\bibinfo {year} {1966})}\BibitemShut {NoStop}%
\bibitem [{\citenamefont {Wang}\ \emph {et~al.}(2016)\citenamefont {Wang}, \citenamefont {Du}, \citenamefont {Liu}, \citenamefont {Hu}, \citenamefont {Zhang}, \citenamefont {Zhang}, \citenamefont {Owen}, \citenamefont {Lu}, \citenamefont {Gan}, \citenamefont {Sengupta}, \citenamefont {Kloc},\ and\ \citenamefont {Xiong}}]{Wang2016}%
  \BibitemOpen
  \bibfield  {author} {\bibinfo {author} {\bibfnamefont {X.}~\bibnamefont {Wang}}, \bibinfo {author} {\bibfnamefont {K.}~\bibnamefont {Du}}, \bibinfo {author} {\bibfnamefont {Y.~Y.~F.}\ \bibnamefont {Liu}}, \bibinfo {author} {\bibfnamefont {P.}~\bibnamefont {Hu}}, \bibinfo {author} {\bibfnamefont {J.}~\bibnamefont {Zhang}}, \bibinfo {author} {\bibfnamefont {Q.}~\bibnamefont {Zhang}}, \bibinfo {author} {\bibfnamefont {M.~H.~S.}\ \bibnamefont {Owen}}, \bibinfo {author} {\bibfnamefont {X.}~\bibnamefont {Lu}}, \bibinfo {author} {\bibfnamefont {C.~K.}\ \bibnamefont {Gan}}, \bibinfo {author} {\bibfnamefont {P.}~\bibnamefont {Sengupta}}, \bibinfo {author} {\bibfnamefont {C.}~\bibnamefont {Kloc}},\ and\ \bibinfo {author} {\bibfnamefont {Q.}~\bibnamefont {Xiong}},\ }\bibfield  {title} {\bibinfo {title} {Raman spectroscopy of atomically thin two-dimensional magnetic iron phosphorus trisulfide ({FePS}\textsubscript{3}) crystals},\ }\href {https://doi.org/10.1088/2053-1583/3/3/031009} {\bibfield  {journal} {\bibinfo
  {journal} {2D Materials}\ }\textbf {\bibinfo {volume} {3}},\ \bibinfo {pages} {031009} (\bibinfo {year} {2016})}\BibitemShut {NoStop}%
\bibitem [{\citenamefont {Rezende}\ \emph {et~al.}(2019)\citenamefont {Rezende}, \citenamefont {Azevedo},\ and\ \citenamefont {Rodríguez-Suárez}}]{Rezende2019}%
  \BibitemOpen
  \bibfield  {author} {\bibinfo {author} {\bibfnamefont {S.~M.}\ \bibnamefont {Rezende}}, \bibinfo {author} {\bibfnamefont {A.}~\bibnamefont {Azevedo}},\ and\ \bibinfo {author} {\bibfnamefont {R.~L.}\ \bibnamefont {Rodríguez-Suárez}},\ }\bibfield  {title} {\bibinfo {title} {Introduction to antiferromagnetic magnons},\ }\href {https://doi.org/10.1063/1.5109132} {\bibfield  {journal} {\bibinfo  {journal} {Journal of Applied Physics}\ }\textbf {\bibinfo {volume} {126}},\ \bibinfo {pages} {151101} (\bibinfo {year} {2019})}\BibitemShut {NoStop}%
\bibitem [{\citenamefont {Coey}(2010)}]{Coey2010}%
  \BibitemOpen
  \bibfield  {author} {\bibinfo {author} {\bibfnamefont {J.~M.~D.}\ \bibnamefont {Coey}},\ }\href {https://doi.org/https://doi.org/10.1017/CBO9780511845000} {\emph {\bibinfo {title} {Magnetism and Magnetic Materials}}}\ (\bibinfo  {publisher} {Cambridge University Press},\ \bibinfo {year} {2010})\BibitemShut {NoStop}%
\bibitem [{\citenamefont {Verma}(1972)}]{Neelmani1972}%
  \BibitemOpen
  \bibfield  {author} {\bibinfo {author} {\bibfnamefont {N.~G.~S.}\ \bibnamefont {Verma}},\ }\bibfield  {title} {\bibinfo {title} {Phonon conductivity of trivalent rare-earth-doped gallium and aluminium garnets},\ }\href {https://doi.org/10.1103/PhysRevB.6.3509} {\bibfield  {journal} {\bibinfo  {journal} {Phys. Rev. B}\ }\textbf {\bibinfo {volume} {6}},\ \bibinfo {pages} {3509} (\bibinfo {year} {1972})}\BibitemShut {NoStop}%
\bibitem [{\citenamefont {Kittel}(2004)}]{Kittel2004}%
  \BibitemOpen
  \bibfield  {author} {\bibinfo {author} {\bibfnamefont {C.}~\bibnamefont {Kittel}},\ }\href {https://www.wiley.com/en-us/Introduction+to+Solid+State+Physics%2C+8th+Edition-p-9780471415268} {\emph {\bibinfo {title} {Introduction to Solid State Physics}}},\ \bibinfo {edition} {eighth edition}\ ed.\ (\bibinfo  {publisher} {John Wiley},\ \bibinfo {year} {2004})\BibitemShut {NoStop}%
\bibitem [{\citenamefont {Rezende}\ \emph {et~al.}(2014{\natexlab{b}})\citenamefont {Rezende}, \citenamefont {Rodr\'{\i}guez-Su\'arez}, \citenamefont {Cunha}, \citenamefont {Rodrigues}, \citenamefont {Machado}, \citenamefont {Fonseca~Guerra}, \citenamefont {Lopez~Ortiz},\ and\ \citenamefont {Azevedo}}]{Rezende2014-2}%
  \BibitemOpen
  \bibfield  {author} {\bibinfo {author} {\bibfnamefont {S.~M.}\ \bibnamefont {Rezende}}, \bibinfo {author} {\bibfnamefont {R.~L.}\ \bibnamefont {Rodr\'{\i}guez-Su\'arez}}, \bibinfo {author} {\bibfnamefont {R.~O.}\ \bibnamefont {Cunha}}, \bibinfo {author} {\bibfnamefont {A.~R.}\ \bibnamefont {Rodrigues}}, \bibinfo {author} {\bibfnamefont {F.~L.~A.}\ \bibnamefont {Machado}}, \bibinfo {author} {\bibfnamefont {G.~A.}\ \bibnamefont {Fonseca~Guerra}}, \bibinfo {author} {\bibfnamefont {J.~C.}\ \bibnamefont {Lopez~Ortiz}},\ and\ \bibinfo {author} {\bibfnamefont {A.}~\bibnamefont {Azevedo}},\ }\bibfield  {title} {\bibinfo {title} {Magnon spin-current theory for the longitudinal spin-seebeck effect},\ }\href {https://doi.org/10.1103/PhysRevB.89.014416} {\bibfield  {journal} {\bibinfo  {journal} {Phys. Rev. B}\ }\textbf {\bibinfo {volume} {89}},\ \bibinfo {pages} {014416} (\bibinfo {year} {2014}{\natexlab{b}})}\BibitemShut {NoStop}%
\bibitem [{\citenamefont {Dolleman}\ \emph {et~al.}(2017)\citenamefont {Dolleman}, \citenamefont {Houri}, \citenamefont {Davidovikj}, \citenamefont {Cartamil-Bueno}, \citenamefont {Blanter}, \citenamefont {van~der Zant},\ and\ \citenamefont {Steeneken}}]{Dolleman2017}%
  \BibitemOpen
  \bibfield  {author} {\bibinfo {author} {\bibfnamefont {R.~J.}\ \bibnamefont {Dolleman}}, \bibinfo {author} {\bibfnamefont {S.}~\bibnamefont {Houri}}, \bibinfo {author} {\bibfnamefont {D.}~\bibnamefont {Davidovikj}}, \bibinfo {author} {\bibfnamefont {S.~J.}\ \bibnamefont {Cartamil-Bueno}}, \bibinfo {author} {\bibfnamefont {Y.~M.}\ \bibnamefont {Blanter}}, \bibinfo {author} {\bibfnamefont {H.~S.~J.}\ \bibnamefont {van~der Zant}},\ and\ \bibinfo {author} {\bibfnamefont {P.~G.}\ \bibnamefont {Steeneken}},\ }\bibfield  {title} {\bibinfo {title} {Optomechanics for thermal characterization of suspended graphene},\ }\href {https://doi.org/10.1103/PhysRevB.96.165421} {\bibfield  {journal} {\bibinfo  {journal} {Phys. Rev. B}\ }\textbf {\bibinfo {volume} {96}},\ \bibinfo {pages} {165421} (\bibinfo {year} {2017})}\BibitemShut {NoStop}%
\bibitem [{\citenamefont {Zhang}\ \emph {et~al.}(2021)\citenamefont {Zhang}, \citenamefont {Nie}, \citenamefont {Wang}, \citenamefont {Xia},\ and\ \citenamefont {Guo}}]{Jianmin2021}%
  \BibitemOpen
  \bibfield  {author} {\bibinfo {author} {\bibfnamefont {J.~M.}\ \bibnamefont {Zhang}}, \bibinfo {author} {\bibfnamefont {Y.~Z.}\ \bibnamefont {Nie}}, \bibinfo {author} {\bibfnamefont {X.~G.}\ \bibnamefont {Wang}}, \bibinfo {author} {\bibfnamefont {Q.~L.}\ \bibnamefont {Xia}},\ and\ \bibinfo {author} {\bibfnamefont {G.~H.}\ \bibnamefont {Guo}},\ }\bibfield  {title} {\bibinfo {title} {Strain modulation of magnetic properties of monolayer and bilayer {FePS}\textsubscript{3} antiferromagnet},\ }\href {https://doi.org/https://doi.org/10.1016/j.jmmm.2020.167687} {\bibfield  {journal} {\bibinfo  {journal} {Journal of Magnetism and Magnetic Materials}\ }\textbf {\bibinfo {volume} {525}},\ \bibinfo {pages} {167687} (\bibinfo {year} {2021})}\BibitemShut {NoStop}%
\bibitem [{\citenamefont {Houtappel}(1950)}]{Houtappel1950}%
  \BibitemOpen
  \bibfield  {author} {\bibinfo {author} {\bibfnamefont {R.}~\bibnamefont {Houtappel}},\ }\bibfield  {title} {\bibinfo {title} {Order-disorder in hexagonal lattices},\ }\href {https://doi.org/https://doi.org/10.1016/0031-8914(50)90130-3} {\bibfield  {journal} {\bibinfo  {journal} {Physica}\ }\textbf {\bibinfo {volume} {16}},\ \bibinfo {pages} {425 } (\bibinfo {year} {1950})}\BibitemShut {NoStop}%
\bibitem [{\citenamefont {Pathria}\ and\ \citenamefont {Beale}(2011)}]{Pathria2011}%
  \BibitemOpen
  \bibfield  {author} {\bibinfo {author} {\bibfnamefont {R.}~\bibnamefont {Pathria}}\ and\ \bibinfo {author} {\bibfnamefont {P.~D.}\ \bibnamefont {Beale}},\ }\href {https://doi.org/10.1016/C2009-0-62310-2} {\emph {\bibinfo {title} {Statistical Mechnics}}}\ (\bibinfo  {publisher} {Elsevier},\ \bibinfo {year} {2011})\BibitemShut {NoStop}%
\bibitem [{\citenamefont {Matveev}\ and\ \citenamefont {Shrock}(1996)}]{Matveev1996}%
  \BibitemOpen
  \bibfield  {author} {\bibinfo {author} {\bibfnamefont {V.}~\bibnamefont {Matveev}}\ and\ \bibinfo {author} {\bibfnamefont {R.}~\bibnamefont {Shrock}},\ }\bibfield  {title} {\bibinfo {title} {Complex-temperature singularities in the $d=2$ ising model: triangular and honeycomb lattices},\ }\href {https://doi.org/10.1088/0305-4470/29/4/009} {\bibfield  {journal} {\bibinfo  {journal} {Journal of Physics A: Mathematical and General}\ }\textbf {\bibinfo {volume} {29}},\ \bibinfo {pages} {803} (\bibinfo {year} {1996})}\BibitemShut {NoStop}%
\bibitem [{\citenamefont {Prasai}\ \emph {et~al.}(2017)\citenamefont {Prasai}, \citenamefont {Trump}, \citenamefont {Marcus}, \citenamefont {Akopyan}, \citenamefont {Huang}, \citenamefont {McQueen},\ and\ \citenamefont {Cohn}}]{Prasai2017}%
  \BibitemOpen
  \bibfield  {author} {\bibinfo {author} {\bibfnamefont {N.}~\bibnamefont {Prasai}}, \bibinfo {author} {\bibfnamefont {B.~A.}\ \bibnamefont {Trump}}, \bibinfo {author} {\bibfnamefont {G.~G.}\ \bibnamefont {Marcus}}, \bibinfo {author} {\bibfnamefont {A.}~\bibnamefont {Akopyan}}, \bibinfo {author} {\bibfnamefont {S.~X.}\ \bibnamefont {Huang}}, \bibinfo {author} {\bibfnamefont {T.~M.}\ \bibnamefont {McQueen}},\ and\ \bibinfo {author} {\bibfnamefont {J.~L.}\ \bibnamefont {Cohn}},\ }\bibfield  {title} {\bibinfo {title} {Ballistic magnon heat conduction and possible poiseuille flow in the helimagnetic insulator {Cu}\textsubscript{2}{OSeO}\textsubscript{3}},\ }\href {https://doi.org/10.1103/PhysRevB.95.224407} {\bibfield  {journal} {\bibinfo  {journal} {Phys. Rev. B}\ }\textbf {\bibinfo {volume} {95}},\ \bibinfo {pages} {224407} (\bibinfo {year} {2017})}\BibitemShut {NoStop}%
\bibitem [{\citenamefont {Argyle}\ \emph {et~al.}(1967)\citenamefont {Argyle}, \citenamefont {Miyata},\ and\ \citenamefont {Schultz}}]{Argyle1967}%
  \BibitemOpen
  \bibfield  {author} {\bibinfo {author} {\bibfnamefont {B.~E.}\ \bibnamefont {Argyle}}, \bibinfo {author} {\bibfnamefont {N.}~\bibnamefont {Miyata}},\ and\ \bibinfo {author} {\bibfnamefont {T.~D.}\ \bibnamefont {Schultz}},\ }\bibfield  {title} {\bibinfo {title} {Magnetoelastic behavior of single-crystal {E}uropium {O}xide. {I}. thermal expansion anomaly},\ }\href {https://doi.org/10.1103/PhysRev.160.413} {\bibfield  {journal} {\bibinfo  {journal} {Phys. Rev.}\ }\textbf {\bibinfo {volume} {160}},\ \bibinfo {pages} {413} (\bibinfo {year} {1967})}\BibitemShut {NoStop}%
\bibitem [{\citenamefont {Shapira}\ and\ \citenamefont {Oliveira}(1978)}]{Shapira1978}%
  \BibitemOpen
  \bibfield  {author} {\bibinfo {author} {\bibfnamefont {Y.}~\bibnamefont {Shapira}}\ and\ \bibinfo {author} {\bibfnamefont {N.~F.}\ \bibnamefont {Oliveira}},\ }\bibfield  {title} {\bibinfo {title} {Magnetostriction, magnetoelastic coupling, and the magnetic {G}r\"uneisen constant in the antiferromagnet {RbMnF}\textsubscript{3}},\ }\href {https://doi.org/10.1103/PhysRevB.18.1425} {\bibfield  {journal} {\bibinfo  {journal} {Phys. Rev. B}\ }\textbf {\bibinfo {volume} {18}},\ \bibinfo {pages} {1425} (\bibinfo {year} {1978})}\BibitemShut {NoStop}%
\bibitem [{\citenamefont {Shapira}\ \emph {et~al.}(1976)\citenamefont {Shapira}, \citenamefont {Yacovitch}, \citenamefont {Becerra}, \citenamefont {Foner}, \citenamefont {McNiff}, \citenamefont {Nelson},\ and\ \citenamefont {Gunther}}]{Shapira1976}%
  \BibitemOpen
  \bibfield  {author} {\bibinfo {author} {\bibfnamefont {Y.}~\bibnamefont {Shapira}}, \bibinfo {author} {\bibfnamefont {R.~D.}\ \bibnamefont {Yacovitch}}, \bibinfo {author} {\bibfnamefont {C.~C.}\ \bibnamefont {Becerra}}, \bibinfo {author} {\bibfnamefont {S.}~\bibnamefont {Foner}}, \bibinfo {author} {\bibfnamefont {E.~J.}\ \bibnamefont {McNiff}}, \bibinfo {author} {\bibfnamefont {D.~R.}\ \bibnamefont {Nelson}},\ and\ \bibinfo {author} {\bibfnamefont {L.}~\bibnamefont {Gunther}},\ }\bibfield  {title} {\bibinfo {title} {Magnetostriction and the two-spin correlation function in {E}u{O}},\ }\href {https://doi.org/10.1103/PhysRevB.14.3007} {\bibfield  {journal} {\bibinfo  {journal} {Phys. Rev. B}\ }\textbf {\bibinfo {volume} {14}},\ \bibinfo {pages} {3007} (\bibinfo {year} {1976})}\BibitemShut {NoStop}%
\bibitem [{\citenamefont {Gomes}\ \emph {et~al.}(2019)\citenamefont {Gomes}, \citenamefont {Squillante}, \citenamefont {Seridonio}, \citenamefont {Ney}, \citenamefont {Lagos},\ and\ \citenamefont {de~Souza}}]{Gomes2019}%
  \BibitemOpen
  \bibfield  {author} {\bibinfo {author} {\bibfnamefont {G.~O.}\ \bibnamefont {Gomes}}, \bibinfo {author} {\bibfnamefont {L.}~\bibnamefont {Squillante}}, \bibinfo {author} {\bibfnamefont {A.~C.}\ \bibnamefont {Seridonio}}, \bibinfo {author} {\bibfnamefont {A.}~\bibnamefont {Ney}}, \bibinfo {author} {\bibfnamefont {R.~E.}\ \bibnamefont {Lagos}},\ and\ \bibinfo {author} {\bibfnamefont {M.}~\bibnamefont {de~Souza}},\ }\bibfield  {title} {\bibinfo {title} {Magnetic {G}r\"uneisen parameter for model systems},\ }\href {https://doi.org/10.1103/PhysRevB.100.054446} {\bibfield  {journal} {\bibinfo  {journal} {Phys. Rev. B}\ }\textbf {\bibinfo {volume} {100}},\ \bibinfo {pages} {054446} (\bibinfo {year} {2019})}\BibitemShut {NoStop}%
\bibitem [{\citenamefont {McWhan}\ \emph {et~al.}(1966)\citenamefont {McWhan}, \citenamefont {Souers},\ and\ \citenamefont {Jura}}]{McWhan1966}%
  \BibitemOpen
  \bibfield  {author} {\bibinfo {author} {\bibfnamefont {D.~B.}\ \bibnamefont {McWhan}}, \bibinfo {author} {\bibfnamefont {P.~C.}\ \bibnamefont {Souers}},\ and\ \bibinfo {author} {\bibfnamefont {G.}~\bibnamefont {Jura}},\ }\bibfield  {title} {\bibinfo {title} {Magnetic and structural properties of europium metal and europium monoxide at high pressure},\ }\href {https://doi.org/10.1103/PhysRev.143.385} {\bibfield  {journal} {\bibinfo  {journal} {Phys. Rev.}\ }\textbf {\bibinfo {volume} {143}},\ \bibinfo {pages} {385} (\bibinfo {year} {1966})}\BibitemShut {NoStop}%
\bibitem [{\citenamefont {Castellanos-Gomez}\ \emph {et~al.}(2013)\citenamefont {Castellanos-Gomez}, \citenamefont {van Leeuwen}, \citenamefont {Buscema}, \citenamefont {van~der Zant}, \citenamefont {Steele},\ and\ \citenamefont {Venstra}}]{Andres2013}%
  \BibitemOpen
  \bibfield  {author} {\bibinfo {author} {\bibfnamefont {A.}~\bibnamefont {Castellanos-Gomez}}, \bibinfo {author} {\bibfnamefont {R.}~\bibnamefont {van Leeuwen}}, \bibinfo {author} {\bibfnamefont {M.}~\bibnamefont {Buscema}}, \bibinfo {author} {\bibfnamefont {H.~S.~J.}\ \bibnamefont {van~der Zant}}, \bibinfo {author} {\bibfnamefont {G.~A.}\ \bibnamefont {Steele}},\ and\ \bibinfo {author} {\bibfnamefont {W.~J.}\ \bibnamefont {Venstra}},\ }\bibfield  {title} {\bibinfo {title} {Single-layer mos2 mechanical resonators},\ }\href {https://doi.org/https://doi.org/10.1002/adma.201303569} {\bibfield  {journal} {\bibinfo  {journal} {Advanced Materials}\ }\textbf {\bibinfo {volume} {25}},\ \bibinfo {pages} {6719} (\bibinfo {year} {2013})}\BibitemShut {NoStop}%
\bibitem [{\citenamefont {Wu}\ \emph {et~al.}(2018)\citenamefont {Wu}, \citenamefont {Liu},\ and\ \citenamefont {Luo}}]{Xufei2018}%
  \BibitemOpen
  \bibfield  {author} {\bibinfo {author} {\bibfnamefont {X.}~\bibnamefont {Wu}}, \bibinfo {author} {\bibfnamefont {Z.}~\bibnamefont {Liu}},\ and\ \bibinfo {author} {\bibfnamefont {T.}~\bibnamefont {Luo}},\ }\bibfield  {title} {\bibinfo {title} {Magnon and phonon dispersion, lifetime, and thermal conductivity of iron from spin-lattice dynamics simulations},\ }\href {https://doi.org/10.1063/1.5020611} {\bibfield  {journal} {\bibinfo  {journal} {Journal of Applied Physics}\ }\textbf {\bibinfo {volume} {123}},\ \bibinfo {pages} {085109} (\bibinfo {year} {2018})}\BibitemShut {NoStop}%
\bibitem [{\citenamefont {Ando}\ \emph {et~al.}(1998)\citenamefont {Ando}, \citenamefont {Takeya}, \citenamefont {Sisson}, \citenamefont {Doettinger}, \citenamefont {Tanaka}, \citenamefont {Feigelson},\ and\ \citenamefont {Kapitulnik}}]{Yoichi1998}%
  \BibitemOpen
  \bibfield  {author} {\bibinfo {author} {\bibfnamefont {Y.}~\bibnamefont {Ando}}, \bibinfo {author} {\bibfnamefont {J.}~\bibnamefont {Takeya}}, \bibinfo {author} {\bibfnamefont {D.~L.}\ \bibnamefont {Sisson}}, \bibinfo {author} {\bibfnamefont {S.~G.}\ \bibnamefont {Doettinger}}, \bibinfo {author} {\bibfnamefont {I.}~\bibnamefont {Tanaka}}, \bibinfo {author} {\bibfnamefont {R.~S.}\ \bibnamefont {Feigelson}},\ and\ \bibinfo {author} {\bibfnamefont {A.}~\bibnamefont {Kapitulnik}},\ }\bibfield  {title} {\bibinfo {title} {Thermal conductivity of the spin-peierls compound {CuGeO}\textsubscript{3}},\ }\href {https://doi.org/10.1103/PhysRevB.58.R2913} {\bibfield  {journal} {\bibinfo  {journal} {Phys. Rev. B}\ }\textbf {\bibinfo {volume} {58}},\ \bibinfo {pages} {R2913} (\bibinfo {year} {1998})}\BibitemShut {NoStop}%
\end{thebibliography}%

\end{document}